\documentclass[a4paper,11pt]{article}

\usepackage{amsmath}
\usepackage{graphicx}
\usepackage{amsfonts}
\usepackage{amssymb}
\usepackage{epsfig}
\usepackage{color}
\usepackage{psfrag}

 \newtheorem{theorem}{Theorem}[section]
    \newtheorem{rema}{Remark}[section]
    
    \newtheorem{propo}[rema]{Proposition}
    
    \newtheorem{defi}[rema]{Definition}
    \newtheorem{lemma}[rema]{Lemma}
    \newtheorem{corol}[rema]{Corollary}

\newcommand{\EQ}{\begin{equation}}
\newcommand{\EN}{\end{equation}}

\newcommand{\sect}[1]{\setcounter{equation}{0}\section{#1}}

\newcommand{\bc}{\begin{center}}
\newcommand{\ec}{\end{center}}
\def\ba#1{\begin{array}{#1}\displaystyle}
\newcommand{\ea}{\end{array}}
\newcommand{\z}{\\[2mm] \displaystyle}

\newcommand{\beq}{\begin{equation}}
\newcommand{\eeq}{\end{equation}}
\newcommand{\beqa}{\begin{eqnarray}}
\newcommand{\eeqa}{\end{eqnarray}}
\newcommand{\no}{\nonumber}
\newcommand{\n}{\nonumber\\}
\newcommand{\bi}{\begin{itemize}}
\newcommand{\ei}{\end{itemize}}

\def\lt#1{\left#1}
\def\rt#1{\right#1}
\def\t#1{\tilde{#1}}

\def\b#1{\bar{#1}}
\def\frc#1#2{\frac{#1}{#2}}

\newcommand{\p}{\partial}

\newcommand{\bra}{\langle}
\newcommand{\ket}{\rangle}

\newcommand{\N}{{\mathbb{N}}}
\newcommand{\R}{{\mathbb{R}}}
\newcommand{\C}{{\mathbb{C}}}
\newcommand{\hC}{{\hat{\mathbb{C}}}}
\newcommand{\uH}{{\mathbb{H}}}
\newcommand{\uL}{{\mathbb{L}}}
\newcommand{\uD}{{\mathbb{D}}}

\newcommand{\Or}{{\cal O}}

\newcommand{\ep}{\epsilon}

\newcommand{\id}{{\rm id}}

\newcommand{\halmos}{\rule{1ex}{1.4ex}}
\newcommand{\eproof}{\hspace*{\fill}\mbox{$\halmos$}}
\newcommand{\proof}{{\em Proof.\ }}

\newcommand{\tto}{\twoheadrightarrow}

\newcommand{\au}{\Xi}

\newcommand{\spa}{{\tt H}}
\newcommand{\spag}{{\tt F}}
\newcommand{\spc}{{\tt C}}

\def\cl#1{\overline{#1}}
\newcommand{\chrg}{\Gamma}
\newcommand{\dd}{{\rm d}}
\newcommand{\bd}{\b{{\rm d}}}
\newcommand{\conn}{\Theta}

\begin{document}

\begin{titlepage}

\begin{center}
{\Large {\bf
Calculus on manifolds of conformal maps and CFT
}

\vspace{1cm}

Benjamin Doyon}

Department of Mathematics, King's College London\\
Strand, London, U.K.\\
email: benjamin.doyon@kcl.ac.uk

\end{center}

\vspace{1cm}

\noindent In conformal field theory (CFT) on simply connected domains of the Riemann sphere, the natural conformal symmetries under self-maps are extended, in a certain way,  to local symmetries under general conformal maps, and this is at the basis of the powerful techniques of CFT. Conformal maps of simply connected domains naturally have the structure of an infinite-dimensional groupoid, which generalizes the finite-dimensional group of self-maps. We put a topological structure on the space of conformal maps on simply connected domains, which makes it into a topological groupoid. Further, we (almost) extend this to a local manifold structure based on the infinite-dimensional Fr\'echet topological vector space of holomorphic functions on a given domain $A$. From this, we develop the notion of conformal $A$-differentiability at the identity. Our main conclusion is that quadratic differentials characterizing cotangent elements on the local manifold enjoy properties similar to those of the holomorphic stress-energy tensor of CFT; these properties underpin the local symmetries of CFT. Applying the general formalism to CFT correlation functions, we show that the stress-energy tensor is exactly such a quadratic differential. This is at the basis of constructing the stress-energy tensor in conformal loop ensembles. It also clarifies the relation between Cardy's boundary conditions for CFT on simply connected domains, and the expression of the stress-energy tensor in terms of metric variations.

\vfill

{\ }\hfill 12 June 2012

\end{titlepage}

\tableofcontents

\sect{Introduction}

Thanks to two-dimensional conformal covariance, correlation functions in conformal field theory (CFT) transform, under conformal maps, in simple ways (see for instance \cite{BPZ,Gins,DFMS97}). Although this conformal symmetry is useful, it is not by itself powerful enough to give rise to the large machinery of CFT and the multitude of nontrivial results. In order to do so, one must consider some locality principles of quantum field theory, which point to the existence of quantum fields whose own correlation functions are holomorphic functions of the position. It is such holomorphic quantum fields like the stress-energy tensor, with their special analytic properties, that form the basis for rigorous algebraic constructions of CFT (for instance, vertex operator algebras and representations \cite{LL04}). This paper is the first part of a work aimed at understanding these infinitesimal ``local conformal symmetries'' through a geometric framework of derivatives with respect to conformal maps. Here we develop some basic notions of first derivatives in general, and show that certain fundamental aspects of CFT on simply connected domains, having to do with the stress-energy tensor, arise from studying such derivatives without the need for an underlying quantum field theory structure. We then apply the general formalism to CFT correlation functions on simply connected domains, where the stress-energy tensor is identified with a conformal derivative.

\subsection{Main idea}

Since correlation functions in CFT are conformally covariant, the only non-trivial variations are those in their conformal moduli space (this is the moduli space of the domain with punctures where local fields insertions are present). This moduli space is finite-dimensional, and one can analyze small moduli parameter variations by taking derivatives. It is well known that moduli parameter derivatives are related to the stress-energy tensor \cite{FS86}. We are interested in extending such ideas to a more general context than CFT; in particular, to Schramm-Loewner evolution (SLE) (see the review for physicists \cite{CarSLE05}) and conformal loop ensembles (CLE) \cite{Sh06,ShW07a,ShW07b}, where random variables are generically supported on extended regions, by opposition to local fields of CFT. In such cases, the conformal moduli space is infinite dimensional, hence one needs an infinite-dimensional analysis. Conformal derivatives provide a general framework for such an analysis. One obtains moduli space variations by making transformations that are conformal on the boundary of the domain and on the support of the random variables or local fields, but singular in some region inside the domain.

Small variations with respect to singular conformal maps were used in \cite{DRC} in the context of identifying the stress-energy tensor in SLE at zero central charge (that is, at the value $\kappa=8/3$ where conformal restriction holds). The idea of \cite{DRC} is as follows. One considers a CFT correlation function of primary fields, say on the Riemann sphere, $\bra \prod_{j} \Or_j(z_j)\ket_{\hC}$, and its image under a map  $g$, given by $\prod_j (\p g(z_j))^{\delta_j}(\overline{\p g(z_j)})^{\t\delta_j} \bra \prod_j \Or_j(g(z_j))\ket_{\hC}$. If $g$ is conformal on $\hC$ (a M\"obius map), then the image under $g$ is equal to the initial correlation function: this is conformal covariance. However, if $g$ is not conformal on $\hC$, then we are making a variation in the moduli space, so the image is different. If we choose
\[
	g_\ep(z) = z + \frc{\ep^2 e^{2i\theta}}{w-z}
\]
then $g_\ep$ is not conformal on $\hC$ (this is essentially a Joukowsky transform). In \cite{DRC}, it was noticed (by a very simple and direct calculation) that the ``derivative''
\[
	\lim_{\ep\to0} \frc{8}{\pi \ep^2} \int_0^{2\pi} d\theta\, e^{-2i\theta} \lt(
		\prod_j \Big(\p g_\ep(z_j)\Big)^{\delta_j}\Big(\overline{\p g_\ep(z_j)}\Big)^{\t\delta_j}
			\lt\bra \prod_j \Or_j(g_\ep(z_j))\rt\ket_{\hC}
		- \lt\bra \prod_{j} \Or_j(z_j)\rt\ket_{\hC} \rt)
\]
exactly reproduces the right-hand side of the conformal Ward identities,
\[
	\bra T(w) \prod_j \Or_j(z_j)\ket_\hC =
		\sum_{j} \lt(\frc{\delta_j}{(w-z_j)^2} + \frc1{w-z_j} \frc{\p}{\p z_j} \rt) \bra \prod_{j=1}^n \Or_j(z_j) \ket_\hC.
\]
This gives a geometric interpretation to the algebraic formula $T(w) = L_{-2}{\bf 1}(w)$ that identifies the holomorphic stress-energy tensor $T(w)$ with a descendent of the identity field ${\bf 1}(w)$ (the pole at $z=w$ in $g_\ep(z)$ corresponds to the application of $L_{-2}$).

The geometric properties of the Joukowsky transform lead to an interpretation of $T(w)$ in the context of SLE \cite{DRC}, and this construction was generalized to (the dilute regime of) CLE \cite{I}, which has a nonzero central charge. We note that the idea of relating small singular conformal transformation to the stress-energy tensor was also discussed in a different way in \cite{F09}.

In the present paper, we put these ideas in a more general and geometric context, generalizing not only to non-primary fields, but also to situations that {\em a priori} lie outside QFT considerations.

\subsection{The construction}

The space of simply connected domains of the Riemann sphere (of hyperbolic type), along with the conformal maps between them, forms a groupoid $\spc$. We put a non-Hausdorff topology on this groupoid that correspond to a certain compact convergence of conformal maps, and that makes maps that are analytic continuation of each other inseparable. Interpreting as a tangent bundle the vector bundle based on $\spc$ where each fiber is the topological space of vector fields on the corresponding domain (isomorphic to the topological space of holomorphic functions), we (almost) arrive at a notion of local Fr\'echet manifold. We study the derivatives on this manifold at the groupoid element given by the identity map on a domain $A$ ({\em conformal $A$-derivatives}). These derivatives are based on the notion of Hadamard derivatives on topological vector spaces. They are elements of the continuous dual of the space of holomorphic vector fields on $A$, which are classes of quadratic differentials (on domains of the Riemann sphere) parametrized by their singularity structure in $A$. In each class one can choose a differential almost holomorphic on the complement $\hC\setminus A$. We show that when there is stationarity under M\"obius maps near to the identity, then it is a holomorphic differential (which we call the {\em global holomorphic $A$-derivative}), and it coordinate-transforms under M\"obius transformations of the coordinates on $\spc$. Further, if there is stationarity under conformal maps on domains complementary to $A$ (essentially, whose exterior does not intersect that of $A$), then the global holomorphic $A$-derivative coordinate-transforms under transformations of coordinates on $\spc$ corresponding to conformal maps on $\hC\setminus A$. When this general theory is applied to CFT, where we take conformal derivatives of correlation functions on simply connected domains, we show that identifying the stress-energy tensor in correlation functions with the conjugation of a global holomorphic derivative reproduces the conformal Ward identities and the boundary conditions of Cardy \cite{C84}. The conjugation involves an object formed by ratio of partition functions, the {\em relative partition function}, which is M\"obius invariant. In particular, the one-point function of the stress-energy tensor is a global holomorphic derivative of the logarithm of the relative partition function. We argue that this is in agreement with the well-known CFT formula relating metric variations of the partition function to one-point averages of the stress-energy tensor. Our main results are expressed in the three theorems of Section \ref{sectConfDiff} and the single theorem of Section \ref{sectWard}.

We note that in the context of the CLE construction \cite{I}, the transformation and analytic properties of the global holomorphic derivative proved here are essential, as well as the fact, also proved here, that the conformal Ward identities in CFT can be expressed using this general derivative concept. It was also in the context of CLE that the one-point average of the stress-energy tensor was first identified with the global holomorphic derivative an object that can be seen as the CLE equivalent of the CFT relative partition function.

\subsection{Relations with some previous works}

Many previous works are related to the present one. However, we have not seen in the literature either the theorems of Section \ref{sectConfDiff} or those of Section \ref{sectWard}; nor have we seen the concept of relative partition function.

{\bf Analytic geometry of CFT}. At the initial stages of the development of CFT in the physics community, there was a strong interest in understanding the geometro-analytic meaning of various aspects of CFT; in particular of the identification of the stress-energy tensor with metric variations. In the important work \cite{FS86}, these metric variations are naturally associated with variations of conformal structures of compact Riemann surfaces with punctures; such variations are described by Beltrami differentials $\mu$, which form the tangent space in this formulation. Equation (6) of \cite{FS86} states that the CFT partition function ${\rm Z}$, which depends on the Riemann surface and on the choice of metric on it, satisfies
\beq\label{formFS}
	\frc{i}{2\pi} \int d z \,d \b{z} \,T(\b{m},m,z) \mu(z,\b{z}) =
	- {\rm Z}^{-1}\, \delta_\mu {\rm Z},
\eeq
where $\delta_\mu {\rm Z}$ is the infinitesimal variation of ${\rm Z}$ in the direction $\mu$, $T(\b{m},m,z)$ is the one-point function of the stress-energy tensor $T(z)$ at the point $z$ on the Riemann surface characterized by the moduli $\b{m},m$, and on both sides a particular metric has been chosen whereby the scalar curvature is constant and the volume is 1 (since $z\mapsto T(\b{m},m,z)$ is not a quadratic differential because of the conformal anomaly, this is important for the left-hand side to be well defined).

By contrast, here, conformal maps rather than conformal structures are the objects that are affected by infinitesimal variations, and such variations are described by holomorphic vector fields $h$ rather than Beltrami differentials. Equation (\ref{formFS}) is to be compared with a formula that follows from Section \ref{sectWard}:
\beq\label{formDCFT}
	\frc1{2\pi i} \int (dz\, h(z) + d\b{z} \,\b{h}(\b{z}))\,
	\bra T(z) \Or({\bf w})\ket = - Z^{-1} \nabla_h (Z \bra \Or({\bf w})\ket),
\eeq
where the correlation function $\bra \cdot\ket$ is evaluated on a simply connected domain $C$, $\Or({\bf w})$ represents a product of fields at non-coincident points $w_1, w_2,\ldots\in C$ away from $z$, the contour of integration on the left-hand side is counter-clockwise in $C$ and does not surround any point $w_i$, $h$ is a holomorphic function on a simply connected domain whose complement is surrounded by the contour of integration, $\nabla_h$ is the associated conformal derivative, and $Z$ is the relative partition function.

In principle, the result (\ref{formDCFT}) could be deduced from an adaptation of (\ref{formFS}) to the context of boundary CFT on simply connected domains with punctures (based on the formal relation $\mu \propto \b\p h$), where the punctures are the points $w_1, w_2,\ldots$. Further, it is likely that both formulations -- variations of conformal structures and variations of conformal maps -- can be connected, through the theory of quasiconformal maps. Developing this connection, however, is beyond the scope of this paper. Note that the use of infinitesimal variations of conformal maps, rather than conformal structures, seems better adapted to the application to SLE and CLE \cite{I}. Note also our use of the relative partition function rather than the partition function, which has nicer transformation properties (M\"obius invariance).

{\bf Geometry of the Virasoro group and Teichm\"uller theory.} 
Differentiable manifolds of conformal maps occur in many situations, and have been related to CFT (and to string theory) in various works in the past.

The Fr\'echet manifold ${\rm Diff}(S^1)/S^1$ of orientation-preserving diffeomorphisms of the circle modulo rigid rotations can be seen as a set of conformal maps through conformal welding \cite{K87,KY87}; also, the universal Teichm\"uller space is a Banach manifold of conformal maps \cite{Lehto}. In fact, it turns out that ${\rm Diff}(S^1)$ has a natural complex-analytic embedding into the universal Teichm\"uller space \cite{NV90}. The universal Teichm\"uller space can further be represented using an open ball in the space of $L^\infty$ functions on the unit disk (the space of Beltrami differentials) \cite{Lehto}. Hence, there are natural relations between variations of Beltrami differentials, variations of certain conformal maps, and variations of elements in ${\rm Diff}(S^1)$, and all these variations have interpretations through infinite-dimensional manifolds. These have been studied extensively in the literature (see e.g.~\cite{BR87,KY87,Z87,N93,GL06}), and in particular in the context of SLE \cite{FrieKal,F08,F09}. Further, the Fr\'echet Lie group ${\rm Diff}(S^1)$ is closely connected to CFT: it has an essentially unique central extension, the Virasoro-Bott group, whose Lie algebra is the Virasoro algebra. Combined with the concept of geometric quantization (or the orbits method in representation theory) one obtains a geometric understanding of CFT -- see for instance \cite{Z87,BR87,T93}.

In these contexts, the sets of conformal maps are proper subsets of the set of conformal maps on a fixed simply connected hyperbolic domain (in particular, there is the requirement that they have a quasi-conformal extension to the plane), and the tangent space can be seen as a proper subset of the set of holomorphic functions on this domain. Technically, our construction differs in two points: (1) our tangent space is the Fr\'echet space of all holomorphic vector fields, and we use the particularities of this space to obtain our precise statements (e.g.~about the dual space); (2) we do not fix the domain: we consider the groupoid of hyperbolic domains, with all conformal maps between them. Hence in particular, the theorems of Section \ref{sectConfDiff} have no immediate equivalent in the context of the universal Teichm\"uller space or of ${\rm Diff}(S^1)$.

Most importantly, the motivation for our construction is the different physical point of view that we take: we consider CFT as a statistical field theory, or a measure theory. The Lie group ${\rm Diff}(S^1)$ or the manifold ${\rm Diff}(S^1)/S^1$ are useful when quantizing a system of field configurations on the circle or when using representations of the Virasoro algebra for an algebraic construction of CFT, but we believe they are not natural when considering random objects on a domain of the Riemann sphere. In particular, ${\rm Diff}(S^1)$ (or its central extension) cannot be the right symmetry group for CFT, as it has a real Virasoro Lie algebra, whereas it is its complexification that is needed. This may be related to the fact that we need Euclidean field theory, Wick rotated as compared to a theory obtained by quantization on the circle.

Conformal maps between domains are the right symmetries, and the structure obtained is that under composition of conformal maps rather than composition of $S^1$-diffeomorphisms. This is a groupoid, a notion which was also introduced in this context in \cite{F09}. It also gives rise to different Lie-theoretic structures than those implied by the ${\rm Diff}(S^1)$ picture. Seeing a simply connected hyperbolic domain as a one-dimensional complex (non-compact) manifold, and in analogy with manifold theory of diffeomorphisms of compact manifolds \cite{A66,H82,M83}, we then think of the tangent space as the space of holomorphic vector fields, whose natural topology is that of compact convergence. It is not natural to consider only the subspace implied by the ${\rm Diff}(S^1)$ picture. Further, for the topology to behave well under compositions of maps, we introduce a certain compact convergence topology on the space of maps where both initial and final domains may vary (see Section \ref{sectFrechMan}). Finally, it is the interplay between conformal variations on domains with disjoint complements that gives the most interesting results. Note in this connection that the Kirillov action of the Virasoro algebra on conformal maps \cite{K98}, as obtained from the ${\rm Diff}(S^1)$ picture, essentially combines conformal variations in two such complementary sectors. Only its complexification contains both types of variations separately. Hence in this sense, our approach provides a way of giving a global structure to this complexification (other ways are the Neretin semigroup \cite{Ner87}, and the finite conformal transformations in the operator formalism of CFT developed in \cite{BB04}).

\subsection{Organization of the paper}

The paper is organized as follows. In Section \ref{sectFrechMan}, we describe the topological groupoid structure of conformal maps, and we overview some ideas as to the generalization of various Lie groups concepts to this groupoid, in order to have a local manifold structure (we do not construct a Lie groupoid structure). In Section \ref{sectConfDiff}, we develop the concept of conformal differentiability in a general setup, and prove the main general theorems of the paper. In Section \ref{sectWard}, we apply the general theory to the case of CFT correlation functions (reviewing their main properties first), and prove the main theorem relating conformal Ward identities and boundary conditions to global holomorphic derivatives. We also provide arguments for the relation with one-point averages of the stress-energy tensor. Finally, in Section \ref{sectConc}, we present our conclusions.

\sect{Groupoid of conformal maps of simply connected domains}\label{sectFrechMan}

Let us consider the set of doublets $\spc=\{(g,A)\}$ where $A$ is any simply connected domain of the Riemann sphere (we will implicitly restrict our attention to the non-trivial cases where $A$ is of hyperbolic type, i.e.~conformal to the disk), and $g:A\to \hC$ is a univalent conformal map (we will denote by $g:A\tto B$ the statement that $g$ is a univalent conformal map of $A$ onto $B$). This space can naturally be given the structure of a groupoid, generalizing that of a group: the product $(g,A)(g',A') = (g\circ g',A')$ is defined if and only if $A = g'(A')$; associativity holds; for every $A$ there is an identity $(\id,A)$; every $(g,A)$ has an inverse $(g^{-1},g(A))$.

Recall that a Lie group is a manifold with a group structure such that the group operations are differentiable maps. Although the related notion of Lie groupoid exists (see \cite{W96} for a nice introduction with examples), it seems to be too restrictive for the construction that we want to develop. Indeed, for the Lie algebroid one considers vector fields that fix the initial domain of the maps (so that one can implement left-invariance); but this does not hold if one follows, by analogy with diffeomorphisms of compact manifold, the usual ways of constructing the exponential map, based on the Fr\'echet manifold of holomorphic vector fields. Hence, we make a slightly different construction, using a non-Hausdorff topology to account for analytic continuations and the changing initial domain, and putting on $\spc$ structures making it possible to define the analogue of a Lie derivative along left-invariant trajectories. We put emphasis on smooth paths rather than on the standard manifold structure. This is in a sense along the lines of a theory of diffeomorphisms of non-compact manifolds -- these are geometrically much more complicated than for compact manifolds (see e.g.~the comments in e.g. \cite{M83}), and it is known that manifolds based on smooth paths are the right objects of have Lie groups of diffeomorphisms of non-compact manifolds (see the comment in the introduction of Chapter IX of \cite{KrieglMichor97}).

The main idea is to describe local deformations around a point $(g,A)$ by the holomorphic vector fields on $A$. We will discuss the following (see for instance the text \cite{V84} for standard Lie group theory):
\begin{itemize}
\item The topological groupoid $\spc$;
\item The vector bundle ${\cal T}\spc$ over $\spc$ where the fibre above $(g,A)$ is the vector space of holomorphic vector fields on the one-complex-dimensional manifold $A$;
\item Left and right actions of $\spc$ on the fibers;
\item Left-invariant local sections and the associated trajectories;
\item Continuous injections of the holomorphic vector fields above $(g,A)$ into a neighborhood of $(g,A)$, constructed using the trajectories of left-invariant local sections.
\end{itemize}
This will lead to a natural differentiability concept giving rise to conformal derivatives.

Note that something similar can be done with nonunivalent conformal functions on the Riemann sphere, interpreted as conformal maps between hyperbolic Riemann surfaces realized as multiple covers of (parts of) the Riemann sphere. We will not need this more general set-up here.

Consider the vector spaces $\spa(A)$ of holomorphic functions on simply connected domains $A$. We will put structures on the groupoid $\spc$ by deriving them from an explicit representation of $\spc$ on $\{\spa(A):A$ simply connected domain$\}$ (this should be compared with defining matrix Lie groups via their fundamental representation). It turns out to be natural to represent $\spc$ via right actions. For all $B\supseteq A$, we define the right action of $(g,B)\in\spc$ on $f\in \spa(A)$ by:
\beqa
	\spa(g(A)) &\to& \spa(A) \n
	f &\mapsto & f\cdot (g,B) := f\circ g \label{actg}
\eeqa
where on the right-hand side, $g$ is restricted to the domain $A$.

\subsection{Topological vector spaces}

We will put on the space $\spa(A)$ of holomorphic functions on $A$ the topology of compact convergence. This topology may be induced from a distance function; for instance, on $\uD$ (the open unit disk), it is given by
\beq\label{dF}
	d_F(h,h') :=
	\sum_{r=1}^\infty 2^{-r} \frc{p_r(h,h')}{1+p_r(h,h')},\quad p_r(h,h') := {\rm sup}(|h(z)-h'(z)|: z\in (1-2^{-r})\uD)
\eeq
for any $h,h'\in\spa(\uD)$. According to this topology, a sequence of conformal maps that converges is one whose maps converge uniformly on any compact subset of $\uD$. This is a Fr\'echet space \cite{Rudin}.

Further, we will denote by $\spa^>(A)$ the space of holomorphic vector fields, or $(-1,0)$-differentials, on $A$. In local coordinates this can be identified with the space of holomorphic functions, and we put the topology on $\spa^>(A)$ induced from that on local-coordinates holomorphic functions (this topology is coordinate independent). If we use the global coordinates $\hC = \C \cup \{\infty\}$ for the Riemann sphere, then the space $\spa^>(A)$ is the space of functions holomorphic on $A-\{\infty\}$, with the condition that if $h\in\spa^>(A)$ and $\infty\in A$, then the function $h(z)/z^2$ has a holomorphic extension to $\infty$. Let $g:A\tto B$. Under this change of coordinates, holomorphic vector fields transform as\footnote{Here and below, juxtaposition means the point-wise product of functions, $\circ$ is the composition and has priority over point-wise product, and $\p$ is the holomorphic derivative; we will also use $\b\p$ for the anti-holomorphic derivative.}
\beq\label{homeo}
	\ba{rcl} {\cal H}_g \;:\; \spa^>(A) & \to & \spa^>(B) \\ h &\mapsto &(h\,\p g)\circ g^{-1}. \ea
\eeq
Note that ${\cal H}_{g_1} {\cal H}_{g_2} = {\cal H}_{g_1\circ g_2}$ for any $g_2:A\tto B$ and $g_1:B\tto C$. Also, for $g:A\tto B$, the map ${\cal H}_g$ is a homeomorphism $\spa^>(A)\to\spa^>(B)$.

Naturally, for all $B\supseteq A$, the action of the vector field $h\in \spa^>(B)$ on $f\in \spa(A)$ is defined by
\beqa
	\spa(A) &\to& \spa(A) \n
	f &\mapsto & f\cdot h := h\p f \label{acth}
\eeqa
where on the right-hand side, $h$ is restricted to the domain $A$.

\subsection{The topological groupoid $\spc$ and the $A$-topology}

The topology on $\spc$ should specialize to that of compact convergence when, in a sequence $(g_n,A_n)$, only the first members $g_n$ of the pairs are considered. Also, this topology should take into consideration that $(g,A)$ and $(g,B)$ act in the same way on $f\in \spa(A)$ if $A\subseteq B$; that is, it should not separate points whose first members are the same, and whose second members are ordered according to the partial order of set inclusion. Given a compact subset $K$ of the simply connected domain $A$ and a number $r>0$, let us defined the open neighborhood $N_{r,K}$ of $(g,A) \in \spc$ by
\[
	N_{r,K} = \big\{(g',A'):{\rm max}\{d(g(z),g'(z)):z\in K\}<r\;,\; A'\supset K\big\}
\]
where $d$ be the distance function for the round metric of the Riemann sphere (we will take it normalized so that its maximum is 1). The topology on $\spc$ is that generated by all such neighborhoods. 

Clearly, this topology is not Hausdorff, since, in agreement with our requirements, if $g$ is conformal on $B$ then the points $(g|_A,A)$ and $(g,B)$ with $A\subset B$ do not have disjoint neighborhoods. Consider the set $N[(g,A)]$ of points of $\spc$ that are in all neighborhoods of $(g,A)$, that is $N[(g,A)]:=\{(g',A'):A\subseteq A',\,g'|_A = g\}$. We will define convergence in $\spc$ in a way that takes into account that $N[(g,A)]\neq (g,A)$ (contrary to a Hausdorff topology): a sequence $(g_n,A_n):n\in \N$ converges to $(g,A)$ if for all neighborhoods $M$ of $(g,A)$, there exists a $m$ such that for all $n>m$, $N[(g_n,A_n)]\cap M\neq\emptyset$. According to this topology, then, a sequence $(g_n,A_n):n\in \N$ converges to $(g,A)$ if and only if:
\begin{enumerate}
\item (see figure \ref{figtopo}) the functions $g_n$ can be conformally and univalently continued to simply connected domains $A_n'\supseteq A_n$, and there exists a sequence $D_1,D_2,\ldots$ of simply connected domains $D_n\subseteq A_n'$ such that $D_n\subseteq D_{n+1}$ and $\lim_{n\to\infty} D_n = A$ (set-theoretically -- that is, $\cup_n D_n = A$);
\begin{figure}
\bc
\includegraphics[width=11cm,height=3cm]{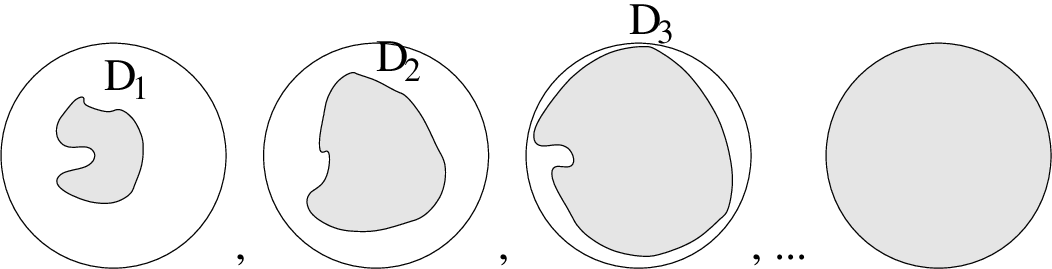}
\ec
\caption{A representation of growing domains where $g_n$ are conformal.}
\label{figtopo}
\end{figure}
\item the sequence of conformally continued functions $g_n:n\in\N$ converges to $g$ compactly on $A$:
\[
	\lim_{n\to\infty} {\rm sup}\{d(g_n(z),g(z)):z\in D_n\} = 0.
\]
\end{enumerate}
We will also say that sequences or families $(g_n,A_n)$ and $(g_n',A_n')$ (over $n$) are inseparable if for every $n$, $(g_n,A_n)$ and $(g_n',A_n')$ do not have disjoint neighborhoods.

Further, we may put a topology on the space of objects (the domains) in the groupoid $\spc$ by simply inducing it from the topology on $\spc$ under the identification $A\leftrightarrow (\id,A)$. This makes $\spc$ into a topological groupoid.

As will become clear below, the domain $A$ in pairs $(g,A)\in\spc$ has two natural meanings: one is that of the domain of the function $g$ (it is from this meaning that the topology on $\spc$ is constructed), the other is that of a choice of coordinates for the space of deformations of $g$. The latter justifies seeing the requirements above for a sequence to converge to $(g,A)$ as a local topology around $(g,A)$, determined by $A$. This will be called the $A$-topology. The $A$-topology is metrizable. Given two points $(g,B)$ and $(g',B')$, let us define the $A$-distance between them as
\beq\label{Adist}
	D_A := \sum_{r=1}^\infty 2^{-r}
		\lt\{\ba{ll} {\rm sup}\{d(g(z),g'(z)):z\in A_r\} & (A_r\subset B\cap B') \\
		1 & (\mbox{otherwise}) \ea \rt.
\eeq
where $A_r:r=1,2,\ldots$ are increasing subsets of $A$ with $\cup_{r=1}^\infty A_r = A$. Then the neighborhoods of $(g,A)$ in the topology induced by $D_A$ are the same as those in the topology on $\spc$.

Finally, we note that it is a simple matter to generalize this to domains with higher connectivity. We will not make use of the general case, except for the local topology around the identity in the case of annular domains (i.e.~doubly connected). This can be derived from the simply connected case as follows, with the use the theorem of Appendix \ref{appfact}. For two simply connected domains $A$ and $B$ such that $\hC\setminus A \subset B$, the set $A\cap B$ is an annular domain (and any annular domain is of this type). Given such $A$ and $B$, the neighborhoods of $(\id,A\cap B)$ are generated by the sets of $(g,C)$ such that, on annular domains $C$, we have $g = g_{A}\circ g_B$ where $g_B$ is in a $B$-neighborhood of $\id$ and $g_{A}$ is in a $A$-neighborhood of $\id$\ \footnote{It may seem more natural to replace $g_A$ by $g_{A'}$ in a $A'$-neighborhood of $\id$, for  $A'=\hC\setminus g_B(\hC\setminus A)$. However, since these are neighborhoods of the identity, this leads to the same definition, and the one given above is more convenient.}. The same local topology is obtained by taking $g = \t{g}_B \circ \t{g}_A$ instead (for $\t{g}_A$ and $\t{g}_B$ in a $A$-neighborhood and a $B$-neighborhood of $\id$, respectively). 

\subsection{The vector bundle over $\spc$ and local sections}

Above each point $(g,A)\in\spc$ we raise the fiber $\spa^>(A)$. That is, we form the vector bundle ${\cal T}\spc$ (analogous to the tangent bundle of Lie groups) defined by the projection $\pi:{\cal T}\spc\to\spc$ with $\pi^{-1}((g,A)) = \spa^>(A)$. We may put on $\spa^>:=\{(h,A):h\in \spa^>(A),\; A$ simply connected domains$\}$ a topology similar to that of $\spc$ (which is the compact convergence topology when we restrict to $\spa^>(A)$ for a fixed $A$), and the full vector bundle structure can be obtained by using local continuous sections that are left- and right-invariant.

Left and right actions of $\spc$ on elements of $\spa^>$ are easy to obtain using the explicit representations (\ref{actg}) and (\ref{acth}). Let $A\subseteq B\subseteq C$ be simply connected domains. For $h\in \spa^>(C)$, we find that $(g,B)\cdot h$ acts on $f\in \spa(g(A))$ as $f\cdot (g,B)\cdot h = h\p g\;\p f\circ g$. On the other hand, let $A\subseteq B$ and $g(B)\subseteq C$. For $h\in \spa^>(C)$, we find that $h\cdot (g,B)$ acts on $f$ as $f\cdot h\cdot (g,B) = h\circ g\;\p f\circ g$. Hence, we will refer to the maps
\beq\label{lract}
	h\mapsto h\p g\quad\mbox{and}\quad h\mapsto h\circ g
\eeq
as left and right actions of $(g,B)$ on $h\in\spa^>(C)$ respectively\footnote{Note that the former does not satisfy the defining property of a left action; its proper understanding is obtained through the application to $f$, as is described.}. Note that the homeomorphism ${\cal H}_g$ is the left action of $g$ followed by the right-action of $g^{-1}$ (but not the inverse, because left and right actions do not commute). Note also that the left action is just the application of the vector field on the acting conformal map.

It is natural to associate to any vector $h\in \spa^>(B)$ above the point $(\id,B)$, the local sections
\beqa
	L_{h,B}&:=&N[\{(g,A)\mapsto h\p g\,|_A: (g,A)\in\spc,\,A\subseteq B\}],\\
	R_{h,B}&:=&N[\{(g,A)\mapsto h\circ g\,|_A:(g,A)\in\spc,\,g(A)\subseteq B\}].\no
\eeqa
Here, the symbol $N$ indicates that we must add elements $(g',A')\in N[(g,A)]$ for elements $(g,A)$ in the set displayed, and associate to them the analytic continuation to $A'$ of the corresponding vector field, if it exists. The local section $L_{h,B}$ is left-invariant, while $R_{h,B}$ is right-invariant. Both local sections are described by linear operators on the original vector $h$, and are continuous; hence they partially complete the vector bundle structure of ${\cal T}\spc$.

\subsection{Trajectories on invariant local sections}\label{ssecttraj}

Following the usual arguments in the context of Lie groups, we construct trajectories associated to left-invariant local sections. Consider a map $T\to \spc: t\mapsto (g_t,A_t)$ where $T\subset \R$ is an open interval containing 0. Denote $A:=A_0$ and $g:=g_0$, and let $h\in \spa^>(A)$. Choosing $T$ small enough and fixing $h$ and $g$, there exists a continuous map $t\mapsto (g_t,A_t)$ such that $g_t$ is compactly differentiable with respect to $t$ on $A_t$, and such that the differential equation
\beq
	\frc{d}{dt} f\cdot g_t = f\cdot g_t \cdot h
\eeq
holds for all $f\in \spa(\cap_{t\in T} g_t(A_t))$. This differential equation is equivalent to
\beq\label{eqman}
	\frc{d}{d t} g_t = h\p g_t,
\eeq
where on the right-hand side of (\ref{eqman}), we have the vector $L_{h,A}((g_t,A_t))$. We may choose the domains $A_t$ to be non-decreasing as $|t|\to0$ in such a way that $\cup_{t\in T} A_t = A$. Having fixed the domains $A_t$, the map $t\mapsto (g_t,A_t)$ is unique.

Let us denote the particular solution where $g_0=\id$ by $(g_t^\id,A_t)$. By left-invariance, we have in general
\beq\label{gtli}
	g_t = g\circ g_t^\id
\eeq
where we must choose $A_t$ such that $g_t^\id(A_t)\subset A$. An explicit solution for $g^\id_t$ is obtained by:
\[
	\int_z^{g_t^\id(z)} \frc{du}{h(u)} = t\;\Leftrightarrow\; g_t^\id(z) = f^{-1}(f(z)+t)
\]
where
\beq\label{hf}
	h(z) = 1/\p f(z).
\eeq
This solution immediately implies that $g_t^\id$ also satisfies the differential equation
\beq\label{eqman2}
	\frc{d}{d t} g_t^\id = h \circ g_t^\id.
\eeq
On the right-hand side, we now have the vector $R_{h,A}((g_t^\id,A_t))$. Hence this is the equation for a trajectory along the right-invariant local section characterized by $h$. That is, an identity-passing trajectory is both left- and right-invariant. A similar statement is of course also true for Lie groups. In combination with (\ref{gtli}), eq.~(\ref{eqman2}) gives in general
\beq\label{eqmanr}
	\frc{d}{d t} g_t = {\cal H}_g(h) \circ g_t.
\eeq
That is, in general, in order to describe a left-invariant trajectory as a right-invariant one, we need to conjugate the vector fields by $g$ before constructing the right-invariant section -- again in analogy with Lie groups.

We may shift back by right action the trajectory described by (\ref{eqman2}) in such a way that it passes by $(\id,g(A))$ at $t=0$, instead of $(\id,A)$: we construct $\t{g}_t^\id = g\circ g^\id_t \circ g^{-1}$, and we see that this is an identity-passing trajectory along both the left- and right-invariant local sections characterized by ${\cal H}_g(h)$. The analogy with Lie groups here is that of a change of coordinates: the conjugation of Lie algebra elements is a Lie algebra isomorphism corresponding to a linear change of coordinates on the tangent space. Hence, in the pairs $(g,A)\in\spc$, the member $A$ may be seen as characterizing the coordinate system on the tangent space.

Equations (\ref{eqman}) and (\ref{eqman2}) imply that
\beq\label{eqcond}
	h\circ g_t^\id = h\p g_t^\id,
\eeq
which is the infinitesimal version of $G_\ep \circ g_t = g_t \circ G_\ep$ for some conformal maps $G_\ep = \id + \ep h + o(\ep)$. The solution also shows that $g_t^\id\circ g_{-t}^\id = \id$ for all $t$ in a neighborhood of 0, hence that
\beq
	g_t^\id \circ g_{t'}^\id = g_{t'}^\id \circ g_t^\id = g_{t+t'}^\id.
\eeq
Despite this, the trajectory passing by the identity, $t\mapsto (g_t^\id,A_t):t\in T$, does not in general form a semigroup in $\spc$, because of the disagreement amongst the domains. Yet, it is inseparable from a semigroup: there exists another trajectory that cannot be topologically separated from $t\mapsto (g_t^\id,A_t)$ and that itself forms a semigroup in $\spc$ (the trajectory $t\mapsto (g_t^\id,B)$ for some $B\subseteq \cap_{t\in T}A_t$). Although in general there is disagreement amongst the domains for $t\mapsto (g_t^\id,A_t)$ to form a semigroup, we expect that it be possible to choose $A_t:t\in T$ in such a way that, for all $t,t',t+t'\in T$, we have $g_{t'}^\id(A_{t'}) = A_t$ if $A_{t'}\subseteq A_{t+t'}$, and $g_{t'}^\id(A_{t+t'}) = A_t$ otherwise. Finally, note that if $T$ can be extended to all of $\R$, then we have a one-parameter subgroup of $\spc$; this, we expect, will only occur if $g_t$ are conformal maps that preserve $A$.

\subsection{Continuous injections and the $A^*$-topology}

Let $H(A)\subset \spa^>(A)$ be a neighborhood of $0$. For every $A$, there exists such a neighborhood such that the map
\beq\label{inj}
	\ba{rcl} \exp_A \;:\; H(A) & \to & \spc \\ h &\mapsto & \exp_A(h) := (g_1^\id,A_1)\ea
\eeq
is well defined, where $t\mapsto (g_t^\id,A_t)$ is the left-invariant trajectory as in Subsection \ref{ssecttraj}. This is the exponential map from the tangent space to $\spc$ at the point $(\id,A)$, and it is a continuous injection.

Certainly, $\exp_A$ cannot be a homeomorphism onto a neighborhood of $(\id,A)$, because it selects specific domains. Let us consider instead the map $N\circ \exp_A:h\mapsto N[\exp_A(h)]$. It is continuous, and it is a ``Hausdorff injection'' (two points, in its image, that are separable, have distinct pre-images). If it maps onto a neighborhood of $(\id,A)$, and if its inverse is continuous, then it can be seen as a ``Hausdorff homeomorphism'', and we may have a structure similar to that of a manifold. 

However, it is likely that $N\circ \exp_A$ does not map onto a neighborhood of $(\id,A)$. Indeed, all maps $g_1^\id$ in the image of $\exp_A$ satisfy the condition (\ref{eqcond}), and it is not obvious that there are neighborhoods of $(\id,A)$ where such an equation holds. More precisely, let
\beq
	S(A) = N[\{(g,B)\in \spc: B\subseteq A,\;g(B)\subseteq A,\;\exists \;h\in\spa^>(A)\;|\;h\circ g = h\p g\}]
\eeq
Then it may be that for all neighborhoods $M$ of $(\id,A)$, there are elements $(g,B)\in M$ such that $(g,B)\not\in S(A)$. Note that this is not surprising: the Lie exponential map in the Lie group theory of diffeomorphisms of compact manifolds is known not to be a homeomorphism \cite{M83,H82}, contrary to the finite-dimensional case. Unfortunately, here we do not have an alternative homeomorphism to obtain a more manifold-like structure.

Yet, if we take on $S(A)$ the topology induced by $\spc$, then $N\circ \exp_A$ is at least a local homeomorphism of $H(A)$ around 0, onto a neighborhood of $(\id,A)$ in $S(A)$ (i.e. the pre-image of a sequence that tends to $(\id,A)$ in $S(A)$ is a sequence that tends to 0). Indeed, take $A=\uD$ for simplicity, and consider some $(g,B)\in S(\uD)$ and the function $q(z) = f(g(z)) - f(z)$ where $f$ is defined in (\ref{hf}). The function $q$ is defined and holomorphic on some domain in $\uD$. Taking its derivative, we find $\p q(z) = \p g(z)/h(g(z)) - 1/h(z) = 0$, hence $q(z) = q$ is a constant, which we can always choose to be $q=1$ by an appropriate choice of the scale of $h$; this defines the pre-image $h$. Inverting, we have $g(z) = f^{-1}(f(z)+1)$, hence $g(z) = g_1^\id(z)$. Choosing $(g,B)$ near enough to $(\id,\uD)$, we find that $f$ is large enough, hence that $h$ is small enough to be in $H(\uD)$. Hence, we may see $H(A)$ as a true tangent space for $S(A)$ at $(\id,A)$. This local topology around $(\id,A)$ will be referred to as $A^*$-topology:
\[
	A^*\mbox{-topology} = A\mbox{-topology}\cap S(A).
\]

The injection (\ref{inj}) can be generalized to one in a neighborhood of $(g,A)$ by left action: $(g\exp)_A(h) := (g\circ g^\id_1,A_1)$. It can also be generalized by right action: $(\exp g)_{g^{-1}(A)}(h) := (g^\id_1\circ g,A_1)$. Note that a combination of a left and right actions gives a change of coordinates: $(g\exp g^{-1})_A(h) = \exp_{g(A)}({\cal H}_g(h))$. Using these and the one-parameter semigroup property discussed in Subsection \ref{ssecttraj}, as well as differentiability concepts for infinite-dimensional topological vector spaces, one can then show particular cases of differentiability of the product and inverse operations in $\spc$, in analogy with the basic property of Lie groups. However, besides a full manifold structure, a more complete description would require much more (e.g.~the analogue of a Baker-Campbell-Hausdorff formula).

\subsection{Lie derivatives and conformal differentiability}

The conformal derivative is simply the Lie derivative on the groupoid of conformal maps associated with left-invariant local sections. However, the property of differentiability itself should be more than the existence of Lie derivatives. On $\spc$, we cannot immediately take that on linear topological spaces, because we do not have a homeomorphism to the tangent space; but restricting to the $A^*$-topology around $(\id,A)$, we may. Hence, given a real function $f$ on $\spc$, we may define $A^*$-differentiability at $(\id,A)$ by the fact that there exists an element $\nabla^A f((\id,A))$ (the differential of $f$ at $(\id,A)$) of the continuous dual $\spa^{>*}(A)$ of $\spa^>(A)$ such that
\beq\label{Astardiff}
	\lim_{h\to0} \frc{f(\exp_A(h)) - f((\id,A)) - \nabla^A f((\id,A)) h}{d_F^{(A)}(h,0)} = 0
\eeq
where $d_F^{(A)}$ is the distance (\ref{dF}) induced by (\ref{homeo}) on $\spa^>(A)$ (this is essentially Fr\'echet differentiability). This of course implies that for all $h\in\spa^>(A)$,
\beq\label{Lieder}
	\lim_{\eta\to0} \frc{f(\exp_A(\eta h)) - f((\id,A))}{\eta} = \nabla^A f((\id,A))(h)
\eeq
where the limit exists locally uniformly around $h$. It is important here that we require that $\nabla^A f((\id,A))$ be not only a linear functional on $\spa^>(A)$, but also a continuous one (which is not automatic in infinite dimension).

Thanks to the local homeomorphism $\exp_A$ around $(\id,A)$ in the $A^*$-topology, it is possible to define continuous paths $((g_\eta,A_\eta):\eta>0)$ lying in $S(A)$ that tend to $\id$ at $\eta\to0$ and that are (right-)differentiable at $\eta=0$. More explicitly, these are all continuous paths tending to $(\id,A)$ as $\eta\to0$ with the property that $g_\eta$ is compactly right-differentiable at $\eta=0$ on $A$. If the local homeomorphism $\exp_A$ were in fact a true homeomorphism (which we haven't proven), then, thanks to continuity of the linear functional $\nabla^A f((\id,A))$, the statement above for (\ref{Lieder}) implies that for any such path $((g_\eta,A_\eta):\eta>0)$, we have
\beq\label{pathder}
	\lim_{\eta\to0} \frc{f((g_\eta,A_\eta)) - f((\id,A))}{\eta} = \nabla^A f((\id,A))(h).
\eeq
for some $h\in\spa^>(A)$.

Conformal differentiability can be seen as an extrapolation of these concepts to the full $A$-topology around $(\id,A)$. Since $\exp_A$ is not necessarily a homeomorphism to a neighborhood of $(\id,A)$, the existence of the continuous dual as in (\ref{Astardiff}) or (\ref{Lieder}) is not quite enough. One way of defining conformal differentiability is to ask additionally for a certain ``Lipshitz continuity''. Given two continuous paths $((g_\eta,A_\eta):\eta>0)$ and $((g_\eta',A_\eta'):\eta>0)$ tending to $(\id,A)$ as $\eta\to0$, let us consider the $A$-distance $D_\eta$ between $(g_\eta,A_\eta)$ and $(g_\eta',A_\eta')$ (see (\ref{Adist})). Then a function $f$ is $A$-differentiable at $(\id,A)$ if there exists a continuous dual $\nabla^A f((\id,A))$ such that (\ref{Lieder}) holds for all $h\in\spa^>(A)$, and if additionally, for any two continuous paths $((g_\eta,A_\eta):\eta>0)$ and $((g_\eta',A_\eta'):\eta>0)$ tending to $(\id,A)$ as $\eta\to0$, we have
\beq
	\lim_{\ep\to0} {\rm sup}\lt\{\frc{f((g_\eta,A_\eta))-f((g_\eta',A_\eta'))}{D_\eta}:\eta\in(0,\ep)\rt\} <\infty.
\eeq
These conditions imply that for any continuous path $((g_\eta,A_\eta):\eta>0)$ tending to $(\id,A)$ as $\eta\to0$ with the property that $g_\eta$ is compactly right-differentiable at $\eta=0$ on $A$, (\ref{pathder}) holds for some $h\in\spa^>(A)$. In fact, requiring that (\ref{pathder}) holds for any such path can be seen as another way of defining $A$-differentiability, weaker that the former (this is essentially Hadamard differentiability). In the rest of this paper, we will use this latter definition (which we will make more explicit in the next section).

\begin{figure}
\bc
\includegraphics[width=14cm,height=2.5cm]{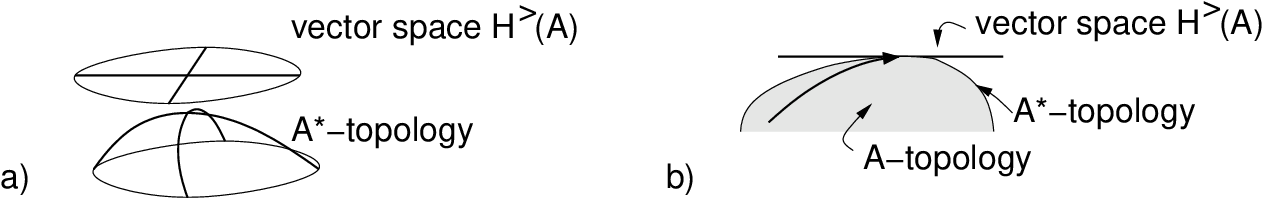}
\ec
\caption{a) A 2-dimensional representation of the topological vector space $\spa^>(A)$ and of the locally homeomorphic $A^*$-topology. Bold lines represent segments of rays on $\spa^>(A)$, and their corresponding paths $g_t$ in the $A^*$-topology. b) The relation between the $A^*$-topology and the $A$-topology, with a path that is compactly differentiable at 0 on $A$.}
\label{figastar}
\end{figure}

Note that $g_\eta$ being compactly right-differentiable at $\eta=0$ on $A$ may be interpreted by saying that the path approaches the identity ``tangentially'' to a ray emanating from the identity by the exponential map. Figure \ref{figastar} gives a (simplistic) pictorial representation of what the relation between the $A$-topology, the $A^*$-topology and the compactly differentiable paths could look like.

\sect{Conformal differentiability} \label{sectConfDiff}

We now make more precise the notion of differentiation that we will use. This notion is based on the geometric ideas of the previous section, but it can be defined essentially independently from them. Notions of differentiability on linear topological spaces are reviewed and studied in \cite{Aver}. Further, calculus on Fr\'echet manifolds is studied quite extensively in \cite{H82}, and in the more general context of the convenient setting (where one only needs a bornology) in \cite{KrieglMichor97}. What we use here is essentially a variant of Hadamard differentiability \cite{Aver}, and we base our definition on smooth paths, as in the ``dynamical'' understanding of the tangent space \cite{KrieglMichor97}. We do not use the definition of continuous differentiability of \cite{H82}, because it is simpler to ask immediately for the existence of a continuous linear functional rather than to derive it from the Fr\'echet manifold structure. Note that certain higher conformal derivatives are studied in \cite{DVOA}, again based on notions of smooth paths (again, as should be natural in view of the comment in the introduction of Chapter IX of \cite{KrieglMichor97}).

In the rest of the paper, for simplicity, we omit the explicit domain $A$ in referring to points in $\spc$; the use of the notion of $A$-topology will guarantee that there is no ambiguity.

\subsection{Paths}

Let $\spag(A)$ be the set of one-parameter families $(g_\eta:\eta>0)$ of maps $A$-converging to the identity $\id$ and compactly right-differentiable at $\eta=0$:
\[
	\spag(A) = \lt\{(g_\eta:\eta>0)\,:\,\ba{ll} \lim_{\eta\to0} g_\eta = \id & \mbox{($A$-topology)} \z
		\lim_{\eta\to0} \frc{G\circ g_\eta\circ G^{-1}-\id}{\eta} & \mbox{exists compactly on $C$} \ea\rt\}
\]
where $C\not\ni \infty$ is a coordinate patch, and $G:A\tto C$ is a M\"obius coordinate map. It is important to note that since we are not considering any particular set of domains associated to $g_\eta$, the restriction that $g_\eta$ be univalent can be lifted: any family of conformal maps $g_\eta:\eta>0$ with $g_\eta\to\id$ compactly on $A$, is such that there exists a family $A_\eta:\eta>0$ of simply connected domains with $\cup_{\eta>0}A_\eta = A$ such that $g_\eta$ is univalent on $A_\eta$. Further, for any $g:A\tto B$, we have
\beq\label{transg}
	\spag(A) = g^{-1}\circ \spag(B)\circ g.
\eeq
Given ${\cal G} = (g_\eta:\eta>0)\in\spag(A)$, the right-derivative of $g_\eta$ at $\eta=0$ is an element of $\spa^>(A)$; we will denote it by $\p {\cal G}$. We find
\beq\label{transgp}
	\p (g\circ {\cal G}\circ g^{-1}) = {\cal H}_g \p{\cal G}.
\eeq

Of course, with ${\cal G}\in \spag(\uD)$, we may simply take $\p{\cal G} := \lim_{\eta\to0} \frc{g_\eta - \id}\eta$. For an explicit general description, we may make use of the global coordinates $\hC$ on the Riemann sphere. Let us denote by $A_\eta\subset A$ domains where $g_\eta$ is conformal, such that $\lim_{\eta\to0}A_\eta = A$. Let us choose some $a \in \hC\setminus A$. If $a=\infty$, let us define $h_\eta^{(a)}$ through
\beq\label{defh1}
	g_\eta(z) = z + \eta h_\eta^{(\infty)}(z).
\eeq
We have that $h_\eta^{(\infty)}$ is holomorphic on $A_\eta$ and converges compactly on $A$ as $\eta\to0$. If $a\neq\infty$, let us define $h_\eta^{(a)}$ through
\beq\label{defh2}
	g_\eta(z) = a + \frc{z-a}{1-\frc{\eta}{z-a}h_\eta^{(a)}(z)}.
\eeq
We have that $h_\eta^{(a)}(z)/(z-a)^2$ is a holomorphic function of $z$ on $A_\eta$ and converges compactly on $A$ as $\eta\to0$. With ${\cal G}\in\spag(A)$, we can define unambiguously
\beq\label{pG2}
	(\p {\cal G})(z) := \lt\{\ba{cl} \displaystyle \lim_{\eta\to0} h_\eta^{(\infty)}(z) & (a=\infty) \z
		(z-a)^2 \lim_{\eta\to0} \frc{h_{\eta}^{(a)}(z)}{(z-a)^2} & (a\neq\infty) \ea\rt.
\eeq
for any $a\in\hC\setminus A$, where in all cases, the limit written exists compactly and is holomorphic for $z\in A$.

Note that for any given $h\in\spa^>(A)$ and any chosen $a\in\hC\setminus A$, we can always form a corresponding family $(g_\eta,\,\eta>0)\in\spag(A)$ by $g_\eta = \exp_A(\eta h)$, or alternatively by
\beq \label{form1}
	g_\eta(z) = z+\eta h(z)\quad (a=\infty)
\eeq
or
\beq \label{form2}
	g_\eta(z) = a + \frc{z-a}{1-\frc{\eta}{(z-a)} h(z)} \quad (a\neq\infty).
\eeq
These families, of course, are different for different $a$, although they lead to the same $h$ (they approach the identity along the same tangent).

\subsection{Continuous duals} \label{ssectcontdual}

Let $\spa^<(A)$ be the vector space of quadratic (i.e. $(2,0)$-) differentials on $A$; these are conjugate to the $(-1,0)$-differentials $\spa^>(A)$ under contour integrals. In local coordinates, $\spa^<(A)$ can of course be identified with the space of holomorphic functions. In global coordinates $\hC$, it is the space of functions holomorphic on $A$, with the requirement that if $u\in\spa^<(A)$ and $\infty\in A$, then $u(z) = O(z^{-4})$ as $z\to\infty$.

According to the general theory (recalled below), for any $\Upsilon\subset \spa^{>*}(A)$, there exists an annular subdomain $U\subset A$ sharing a boundary component with $A$ (which we will refer to as an {\em annular neighborhood of $\p A$ inside $A$}), and a quadratic differential $\gamma\in \spa^<(U)$, such that for every $h\in \spa^{>}(A)$,
\beq\label{Upsilon}
	\Upsilon h = \oint_{z:\vec\p A^-}\dd z\,\gamma(z)h(z) + \oint_{z:\vec\p A^-} \bd\b{z}\,\b\gamma(\b{z})\b{h}(\b{z}).
\eeq
The notation ${z:\vec\p A^-}$ means that the (rectifiable, closed) contour lies in $U$ and goes once in a positive direction around the interior of $A$. Here and below, we normalize the complex integration measure by $\oint \dd z/z = \oint \bd\b{z}/\b{z} = 1$. 

Naturally, we can add to $\gamma$ any quadratic differential on $A$ without changing the result. Hence, a more unique way of characterizing the linear functional is by giving an element of the quotient space $\spa^<(U)/\spa^<(A)$: a class of functions $\{\gamma+u:u\in\spa^<(A)\}$. Further, we may wish to choose particular representatives of these classes. Let $A=\uD$, and consider the global coordinates $\hC$, whereby $\spa^<(\uD)\cong \spa(\uD)$ canonically. By Cauchy's integral formula, given a $\gamma$ in the class, we can write in a unique way $\gamma = u+v$ where $u\in \spa(\uD)$ and $v\in \spa(\hC\setminus \uD)$ with $v(\infty)=0$. Since $\gamma$ is a quadratic differential on $U$, we see that $v$ is in fact a quadratic differential on $\hC\setminus \uD$ except possibly for a singularity of order 3 at $\infty$. We will denote by $\spa_a^<(A)$ the space of quadratic differentials on $A$ with a singularity of maximal order 3 at $a\in A$. In global coordinates, these are functions that are holomorphic on $A$ except for a pole of maximal order 3 at $a$ if $a\neq\infty$, with a behavior $O(z^{-4})$ as $z\to\infty$ if $\infty\in A$ and $a\neq \infty$ or $O(z^{-1})$ if $a=\infty$. Reverting to the abstract Riemann sphere, these global-coordinate arguments imply that for any class $\{\gamma+u:u\in\spa^<(A)\}$ and for any $a\in \hC\setminus A$, there is a unique member of the class that is in $\spa_a^<(\hC\setminus A)$. Hence, since integrals as in (\ref{Upsilon}) always represent continuous linear functionals, we have
\beq\label{dual}
	\spa^{>*}(A) \cong \spa_a^<(\hC\setminus A).
\eeq
A particularly important subspace, which we will denote by $\spa^{>*}_G(A)\subset \spa^{>*}(A)$, is that which is perpendicular to (i.e.~annihilates) the subspace of global vector fields (vector fields on $\hC$). Clearly in this case, the unique member of $\spa_a^<(\hC\setminus A)$ characterizing the functional must have no singularity at the point $a$ (hence it is the same for any $a$), and we have
\beq\label{dualglob}
	\spa^{>*}_\perp(A) \cong \spa^<(\hC\setminus A).
\eeq

Further, we can define a process of continuation along the spaces $\spa_a^<(A)$. Given simply connected domains $A$ and $A'$ such that $A\cap A'$ is simply connected and that $a,a'\in A\cap A'$, we will say that $\gamma \simeq \gamma'$ if $\gamma\in \spa_a(A)$, $\gamma'\in \spa_{a'}(A')$ and there exists a $u\in \spa^<(\hC\setminus (A\cap A'))$ such that $\gamma|_{A\cap A'} = \gamma'|_{A\cap A'} + u|_{A\cap A'}$ (where $u|_{A\cap A'}$ is the analytic continuation of $u$ to $A\cap A'$). This $u$ is of course unique, and can indeed be analytically continued to $A\cap A'$, where it has two maximal-order-3 singularities, one at $a$ the other at $a'$. Moreover, given such $\gamma$, $a$, $A$, $a'$ and $A'$, there is a unique $\gamma'$ such that $\gamma\simeq \gamma'$. For general $a$, $A$, $a'$ and $A'$, we then say that $\gamma\simeq \gamma'$ if the congruence holds as above along a chain that connects $a$, $A$ to $a'$, $A'$. If the quadratic differential $\gamma$ is non-singular, then this process is the usual analytic continuation (of quadratic differentials). In (\ref{Upsilon}), we can deform the integration contour (hence the domain $A$) if we deform accordingly $\gamma\in \spa_a^<(\hC\setminus A)$ following this congruence, as long as we stay in a region of holomorphy of the holomorphic vector field $h$.

The main lemma leading to (\ref{Upsilon}) is as follows. Let us start with $\spa^{>*}(\uD)$ and use global coordinates. In such coordinates, this is canonically $\spa^*(\uD)$, the continuous dual of $\spa(\uD)$. The monomials
\beq
	H_{n,s}(z) = e^{i\pi s/4} z^{n},\,n=0,1,2,\ldots,\,s=\pm
\eeq
form a basis in $\spa(\uD)$: any function $h\in\spa(\uD)$ can be written as a convergent series
\beq
	h = \sum_{n\ge0,s=\pm} c_{n,s}(h) H_{n,s},
\eeq
and the linear functionals $c_{n,s}$ are continuous, since they are given by
\beq\label{coeffexp}
	c_{n,s}(h) =  {\rm Re}\lt[ \oint \dd z\,z^{-n-1} e^{-i\pi s/4} h(z)\rt].
\eeq
Here, the contour lies in $\uD$ and surrounds the point 0 once counter-clockwise.
\begin{lemma}\label{lemlf} (see, for instance, \cite{Conway,Rudin}). For the space $\spa^*(\uD)$ of continuous linear functionals on $\spa(\uD)$, we have:
\begin{itemize}
\item[(a)] Any $\Upsilon\in\spa^*(\uD)$ is completely characterized by the sequence $\{\Upsilon H_{n,s}:n=0,1,2,\ldots,\,s=\pm\}$, in such a way that for any $h\in\spa(\uD)$, we have the convergent series
\beq\label{lflincomb}
	\Upsilon h = \sum_{n\ge 0,s=\pm} c_{n,s}(h) \Upsilon H_{n,s}.
\eeq
\item[(b)] Any $\Upsilon\in\spa^*(\uD)$ is such that
\beq\label{Cz}
	\gamma(z) := \frc12 \sum_{n\ge0,s=\pm} z^{-n-1} e^{-i\pi s/4} \Upsilon H_{n,s}
\eeq
defines a function of $z$ that is holomorphic on $\hC\setminus\uD$.
\item[(c)] Any $\Upsilon\in\spa^*(\uD)$ is completely characterized by the class of functions
\beq\label{claC}
	{\cal C} := \lt\{\gamma+u : u\in\spa\rt\}
\eeq
where $\gamma$ is given by (\ref{Cz}), in such a way that for any $h\in\spa(\uD)$, we have
\beq\label{lfint}
	\Upsilon h = \int_{z:\vec\p\uD^-} \dd z \,\alpha(z)\, h(z) +
	\int_{z:\vec\p\uD^-} \bd \b{z} \,\b\beta(\b{z})\, \b{h}(\b{z})\quad\forall\quad
    \alpha,\beta \in {\cal C}.
\eeq
The function defined by (\ref{Cz}) is the unique member of the class ${\cal C}$ that is holomorphic on $\hC\setminus\uD$ and that vanishes at $\infty$. If (\ref{lfint}) holds for some given $\alpha,\beta$ holomorphic on an annular neighborhood of $\p\uD$ inside $\uD$, and for all $h\in\spa(\uD)$, then it must be that $\alpha,\beta\in{\cal C}$.
\item[(d)] In the sense of (a), the set $\spa^*(\uD)$ is the set of all sequences $\{b_{n,s}\in\R:n=0,1,2,\ldots,\,s=\pm\}$ such that
\beq\label{seriesconv}
	\sum_{n\ge 0,s=\pm} \lt| c_{n,s}(h)\, b_{n,s} \rt| \mbox{ converges }\forall\ h\in\spa.
\eeq
\item[(e)] In the sense of (c), the set $\spa^*(\uD)$ is the set of all classes $\lt\{\gamma+u: u\in\spa\rt\}$ such that $\gamma$ is holomorphic on an annular neighborhood of $\p\uD$ inside $\uD$.
\end{itemize}
\end{lemma}
Elementary proofs of all these statements are presented in Appendix \ref{appproof} (other proofs can be found in, for instance, \cite{Conway,Rudin}).

For the general case $\spa^{>*}(A)$, we may use conformal transport to generalize points (c) and (e) above: by the homeomorphism (\ref{homeo}), we can always write a continuous linear functional $\Upsilon^A$ on $\spa^>(A)$ as $\Upsilon^A = \Upsilon {\cal H}_g^{-1}$ for some $g:\uD\tto A$ and some $\Upsilon\in\spa^{>*}(\uD)$. Then, the discussion above immediately leads to the following.
\begin{lemma}\label{lemlfA} With $A$ a simply connected domain, for the space $\spa^{>*}(A)$ of continuous linear functionals on $\spa^>(A)$, we have:
\begin{itemize}
\item[(a)] Any $\Upsilon^A\in\spa^{>*}(A)$ is completely characterized by the class
\beq\label{claCA}
	{\cal C}^A := \lt\{\gamma+u : u\in\spa^<(A)\rt\}
\eeq
with $\gamma\in\spa^<(U)$ for some annular neighborhood $U$ of $\p A$ inside $A$, in such a way that for any $h\in\spa$, we have
\beq\label{lfintA}
	\Upsilon^A h = \int_{z:\vec\p A^-} \dd z \,\alpha(z)\, h(z) +
	\int_{z:\vec\p A^-} \bd \b{z} \,\b\beta(\b{z})\, \b{h}(\b{z})\quad\forall\quad
    \alpha,\beta \in {\cal C}^A.
\eeq
The function $\gamma$ in (\ref{claCA}) can be chosen, for any given $a\in\hC\setminus A$, as the unique member of ${\cal C}^A$ that is in $\spa_a^<(\hC\setminus A)$. Moreover, if (\ref{lfintA}) holds for some given $\alpha,\beta\in\spa^<(U)$, and for all $h\in\spa^>(A)$, then it must be that $\alpha,\beta\in{\cal C}^A$.
\item[(b)] In the sense of (a), the set $\spa^{>*}(A)$ is the set of all classes $\lt\{\gamma+u: u\in\spa^<(A)\rt\}$ such that $\gamma\in\spa^<(U)$ for some annular neighborhood $U$ of $\p A$ inside $A$.
\end{itemize}
\end{lemma}

A simple consequence of Lemma \ref{lemlfA} is the following quadratic-differential transformation property of the classes characterizing continuous linear functionals.
\begin{lemma}\label{lemtranslf}
With $A$ and $B$ simply connected domains, if ${\cal C}^B$ is the class characterizing the functional $\Upsilon^B\in\spa^{>*}(B)$, then
\beq
	{\cal C}^A = (\p g)^2\, ({\cal C}^B\circ g) \equiv \{(\p g)^2 (\alpha\circ g) : \alpha\in {\cal C}^B\}
\eeq
is the class characterizing $\Upsilon^A = \Upsilon^B {\cal H}_g$, for any conformal $g:A\tto B$.
\end{lemma}

\subsection{Definition of the conformal derivative}

In order to simplify later applications, we will consider, instead of functions on subsets of $\spc$, functions on some abstract set $\Omega$ on which there is an action of $g\in\spc$ in a $A$-neighborhood of the identity (for some simply or doubly connected domain $A$). We will denote by $g\cdot \Sigma$, for $g$ in a $A$-neighborhood of $\id$, the action of $g$ on the point $\Sigma\in\Omega$. This defines a neighborhood of $\Sigma$ in $\Omega$, which we will also call $A$-neighborhood. Throughout, the symbols $\Omega$ and $\Sigma$ (as well as $\Sigma'$, etc.) will be used with this meaning. We may think of $\Sigma$ as being, for instance, a closed subset of a domain of $\hC$. For $g$ in a $A$-neighborhood, the action must satisfy two properties: $\id\cdot\Sigma = \Sigma$, and if $g$ acts on $\Sigma$, then $g'\cdot\Sigma = g\cdot\Sigma$ for all $g'\in N[g]$. Further, given $A$ and $A'$ simply or doubly-connected domains, a $A$-neighborhood $N_{A}(\Sigma)$ of a point $\Sigma\in\Omega$ and a $A'$-neighborhood $N_{A'}(\Sigma')$ of another point $\Sigma'\in\Omega$ will be said to be {\em connected by $g$} for a $g:A\tto A'$, if there exists a bijective map $g\;\cdot: N_A(\Sigma) \to N_{A'}(\Sigma')$, with $\Sigma' = g\cdot \Sigma$, such that for any $\t{g}'$ in a $A'$-neighborhood of $\id$, we have
\beq\label{connected}
	g^{-1}\cdot \t{g}'\cdot g\cdot \Sigma = (g^{-1}\circ \t{g}' \circ g)\cdot \Sigma.
\eeq
The condition in this definition immediately implies that for any $\t{g}$ in a $A$-neighborhood of $\id$, we have $g\cdot \t{g}\cdot g^{-1}\cdot \Sigma' = (g\circ \t{g}\circ g^{-1})\cdot \Sigma'$. Note that we will not need any more properties of actions of maps in $\spc$ than those stated here.

We will study differentiability at $\Sigma$ of $\R$-valued functions $f$ on a $A$-neighborhood in $\Omega$. The restriction to $\R$-valued functions is for simplicity, and also because the applications to probability functions that occur in the context of CLE involve such real functions (this can easily be generalized, for instance, to any normed $\R$- or $\C$-linear space). Below, when we talk about functions without more specification, we will think of $\R$-valued functions on $\Omega$.

\begin{defi}\label{defider}
Let $A$ be a simply connected domain. An $\R$-valued function $f$ on a $A$-neighborhood of $\Sigma$ in $\Omega$ is $A$-differentiable at $\Sigma$ if there exists a continuous linear functional $\nabla^A f(\Sigma)$ on $\spa^>(A)$ such that the following limit exists and gives
\beq\label{derlf}
	\lim_{\eta\to 0} \frc{f(g_\eta\cdot \Sigma) - f(\Sigma)}{\eta} = \nabla^A f(\Sigma) h
\eeq
for any $(g_\eta:\eta>0)\in\spag(A)$, where $h=\p(g_\eta:\eta>0)$.
\end{defi}
In parallel with the usual terminology, we will call $\nabla^A f(\Sigma)$ the {\em conformal derivative} or {\em differential of $f$ at $\Sigma$}, and $\nabla^A f(\Sigma) h$ the {\em directional derivative of $f$ at $\Sigma$ in the direction $h$}. For convenience, we will denote by
\beq
	\nabla_h f(\Sigma) := \nabla^A f(\Sigma) h
\eeq
the directional derivative. In this notation, $\nabla_h\cdot (\Sigma)$ can be seen as an element of the tangent space at $\Sigma$. Clearly, our notation suggests that there may be a real function $\nabla_h f$ on $\Omega$, and a map $\nabla^A f$ from $\Omega$ to continuous linear functionals on $\spa^>(A)$; however, for our purposes it will mostly be sufficient to fix $\Sigma$. Note that the notation $\nabla_h f(\Sigma)$ suggests that this is independent of $A$, and only depends on $h$ (for given $f$ and $\Sigma$); this is very natural, and we will show that it is indeed the case.

From Lemma \ref{lemlf}, we have, in the case where $A=\uD$, that an $\R$-valued function $f$ is $\uD$-differentiable at $\Sigma\in\Omega$ if and only if
\beq\label{derivative}
	f_{n,s}(\Sigma) := \lim_{\eta\to0} \frc{f((\id + \eta H_{n,s})\cdot \Sigma)-f(\Sigma)}\eta \quad \mbox{exists}
\eeq
for all $n=0,1,2,3,\ldots$ and $s=\pm$, and
\beq\label{derseries}
	\lim_{\eta\to0} \frc{f(g_\eta\cdot \Sigma)-f(\Sigma)}\eta
	= \sum_{n\ge0,s=\pm} c_{n,s}(h) f_{n,s}(\Sigma) \quad \mbox{converges}
\eeq
for any $\{g_\eta,\,\eta>0\}\in\spag(\uD)$, where $h=\p\{g_\eta:\eta>0\}$. We may refer to the numbers $f_{n,s}(\Sigma)$ as the {\em partial derivatives of $f$ at $\Sigma$}.

From Lemma \ref{lemlfA}, we also have:
\begin{corol}\label{corolint}
An $\R$-valued function $f$ is $A$-differentiable at $\Sigma\in\Omega$ if and only if there exists a class
\beq
	\Delta^A f(\Sigma) := \{\gamma+u:u\in\spa^<(A)\}
\eeq
where $\gamma\in \spa^<(U)$ for an annular neighborhood $U$ of $\p A$ inside $A$, such that
\beq\label{derint}
	\lim_{\eta\to0} \frc{f(g_\eta\cdot \Sigma)-f(\Sigma)}\eta
	= \int_{z:\vec\p A^-} \dd z\,\alpha(z)\,h(z) + \int_{z:\vec\p A^-} \bd\b{z}\,\b\beta(\b{z})\,\b{h}(\b{z})\quad\forall
	\quad \alpha,\beta \in \Delta^A f(\Sigma)
\eeq
for any $(g_\eta,\,\eta>0)\in\spag(A)$, where $h=\p(g_\eta:\eta>0)$.
\end{corol}
The class $\Delta^A f(\Sigma)$ will be referred to as the {\em holomorphic $A$-class of $f$ at $\Sigma$}. For any $a\in \hC\setminus A$, there is a unique member of this class given by
\beq\label{DeltaA}
	\{z\mapsto \Delta_{a;z}^A f(\Sigma)\} \in \spa_a^<(\hC\setminus A)
\eeq
(here, $z$ should be seen as a point on the abstract Riemann sphere). These will be called {\em holomorphic $A$-derivatives of $f$ at $\Sigma$} (and their complex conjugates $\b\Delta_{a;\b{z}}^A f(\Sigma) := \overline{\Delta_{a;z}^A f(\Sigma)}$, anti-holomorphic $A$-derivatives). In the case $A=\uD$ and $a=\infty$, we simply have, in global coordinates,
\beq\label{DeltauD}
	\Delta_{\infty;z}^\uD f(\Sigma) = \frc12 \sum_{n\ge0,s=\pm} z^{-n-1} e^{-i\pi s/4} f_{n,s}.
\eeq
 
Note that (\ref{derint}) has an intuitive interpretation: it gives us the $A$-derivative in the direction $h$ in a form where $h$ is essentially integrated along $\p A$, as if we were ``summing'' over small contributions from derivatives with respect to all points of the boundary of the domain $A$.

\subsection{General properties}

We first consider properties under change of coordinates on a neighborhood of $\Sigma$:
\begin{propo}\label{propog}
Let $A$ be a simply connected domain and $g:A\tto A'$ a map connecting a $A$-neighborhood of $\Sigma$ to a $A'$-neighborhood of $\Sigma' = g\cdot \Sigma$. Let $f$ be a function on the $A$-neighborhood of $\Sigma$, and define $f' := f\circ g^{-1}$. If $f'$ is $A'$-differentiable at $\Sigma'$, then $f$ is $A$-differentiable at $\Sigma$, and
\beq\label{transA}
	\Delta^A f(\Sigma) = (\p g)^2 \lt(\Delta^{A'} f' (\Sigma')\rt)\circ g.
\eeq
\end{propo}
\proof From (\ref{transg}), (\ref{transgp}) and (\ref{connected}), we have, for $(g_\eta:\eta>0)\in\spag(A)$,
\beqa
	\lim_{\eta\to0} \frc{(f'\circ g)(g_\eta\cdot g^{-1}\cdot \Sigma'))) - (f'\circ g)(g^{-1}\cdot \Sigma')}{\eta} &=&
	\lim_{\eta\to0} \frc{f'((g\circ g_\eta\circ g^{-1})\cdot \Sigma') - f'(\Sigma')}{\eta} \n
	&=& \nabla f'(\Sigma') {\cal H}_g h \no
\eeqa
so that we find differentiability, with $\nabla f(\Sigma) = \nabla f'(\Sigma') {\cal H}_g$. With Lemma \ref{lemtranslf}, this completes the proof. \eproof

Hence, the holomorphic $A$-class transforms like a quadratic differential. This transformation property is purely a class property, and in fact, generically, no member function of this class, ``fixed'' in some way, transforms like this. However, since the holomorphic $A$-derivative is a (singular) quadratic differential on a complement domain, it does have a simple transformation property under M\"obius maps $g=G$:
\beq\label{transmob}
	\Delta_{a;z}^A f(\Sigma) = \p G(z)^2 \Delta^{A'}_{G(a);G(z)} f'(\Sigma').
\eeq
Note that in general, the position of the singularity changes.

We next address the question of the independence upon $A$ of the directional derivative $\nabla_h f(\Sigma) = \nabla^A f(\Sigma) \, h$.
\begin{propo}\label{propoindep}
Consider a function $h\in\spa^>(A)\cap \spa^>(B)$ for two simply connected domains $A$ and $B$ with $A\cap B\neq \emptyset$. If $f$ is both $A$-differentiable and $B$-differentiable at $\Sigma$, then we have
\beq \label{indepint}
	\int_{z:\vec\p A^-} \dd z \,\alpha^A(z)\,h(z) = \int_{z: \vec\p B^-} \dd z \,\alpha^B(z)\,h(z)\quad \forall\quad 
	\alpha^A \in \Delta^A f(\Sigma),\  \alpha^B \in \Delta^B f(\Sigma)
\eeq
so that in particular
\beq \label{indeplf}
	\nabla^A f(\Sigma)\, h = \nabla^B f(\Sigma)\, h.
\eeq
\end{propo}
\proof First, let us consider the case where the complements of $A$ and $B$ have a non-empty intersection, $\hC\setminus A\, \cap \,\hC\setminus B \neq \emptyset$. Let us choose a point $a\in\hC$ that is not in $A\cup B$. Then, we can form the family ${\cal G} = (g_\eta:\eta>0)\in\spag(A)\cap \spag(B)$ such that $h=\p{\cal G}$ by using (\ref{form1}) or (\ref{form2}) as appropriate (depending on $a$). From Definition \ref{defider}, we can write two relations like (\ref{derlf}) for exactly the same limit (the same left-hand side), using $A$-differentiability and $B$-differentiability, so that we obtain (\ref{indeplf}). Moreover, from Corollary \ref{corolint}, we can also write two relations like (\ref{derint}) for the same limit, and repeat the process with the replacement $h\mapsto ih$. Taking linear combinations in order to isolate the holomorphic part, we obtain (\ref{indepint}).

Now let us consider the case where the complements of $A$ and $B$ have empty intersection. Then, the space $\spa^>(A)\cap \spa^>(B)$ is in fact $\spa^>(\hC)$, the six-dimensional space of functions of the form (in global coordinates) $h(z) = a + bz + cz^2,\;a,b,c\in\C$. For any such $h=\p {\cal G}$ we can form the family ${\cal G} = \{g_\eta:\eta>0\}$ of global conformal transformations $g_\eta(z) = ((1+\eta b)z + \eta a)/(1-\eta c z)$. This family is in $\spag(C)$ for any simply connected domain $C$, in particular for $C=A$ and $C=B$. Hence, by the same reasoning as above, we obtain (\ref{indepint}) and (\ref{indeplf}).
\eproof

It is important to realize that in (\ref{indepint}), generically, we are not merely making a change of the integration contour: we are at the same time changing the function that is being integrated, since in general $\alpha^A(z)$ and $\alpha^B(z)$ have different singularity structures outside of $A\cap B$.

It is instructive to look at some simple examples of holomorphic $A$-derivatives. In the case $A = \hC\setminus\cl\uD$, we may use the transformation property (\ref{transA}) with $G(z) = 1/z$, as well as the expression (\ref{DeltauD}). Let us introduce the functions $H_{n,s}(z) = e^{i\pi s/4} z^n$ for integers $n<0$, as well as the corresponding negative-index partial derivatives
\beq
	f_{n,s} = \lim_{\eta\to0} \frc{f((\id + \eta H_{n,s})\cdot \Sigma)-f(\Sigma)}\eta
\eeq
which exist for all $n\leq 2$ if $f$ is $\hC\setminus\cl\uD$-differentiable. Writing $G\circ (\id + \eta H_{n,s}) \circ G = \id - \eta H_{2-n,s}+O(\eta^2)$, and after a shift and change of sign of $n$, we obtain
\beq \label{DeltaCuD}
	\Delta_{0;z}^{\hC\setminus\b\uD} f(\Sigma) =
		-\frc1{2} \sum_{n\le2,\,s=\pm} z^{-n-1} e^{-i\pi s/4} f_{n,s}.
\eeq
Note that it is holomorphic on $\uD$ except for a pole of order 3 at $z=0$. In an entirely similar way, using the scale transformation $g(z) = rz$ for real $r>0$, as well as a re-scaling of the parameter $\eta$, we obtain the following formulae:
\beq\label{Deltascale}
	\Delta_{\infty;z}^{r\uD} f(\Sigma) = \Delta_{\infty;z}^\uD f(\Sigma),\quad
	 \Delta_{0;z}^{\hC\setminus r\cl\uD} f(\Sigma) = \Delta_{0;z}^{\hC\setminus \cl\uD} f(\Sigma).
\eeq

These equalities have a generalization: a theorem that allows us to change the domain of differentiability. It is based on the idea that if $f$ is $A$-differentiable at $\Sigma$, then it should also be $B$-differentiable at $\Sigma$ for any $B$ such that $A\subseteq B$, because small conformal transformations on $B$ necessarily produce small conformal transformations on $A$. This is true, and the following proposition gives us also the relation between the holomorphic derivatives for different domains of differentiability, for fixed $f$ and $\Sigma$.

\begin{propo}\label{propoAB}
If a function $f$ is $A$-differentiable at $\Sigma$ for some simply connected domain $A$, then it is also $B$-differentiable at $\Sigma$ for any simply connected domain $B\supseteq A$. Moreover we have, for any $a\in \hC\setminus B$,
\beq
	\Delta_{a;z}^A f(\Sigma) = \Delta_{a;z}^B f(\Sigma).
\eeq
\end{propo}
\proof Let us consider $\nabla_h f(\Sigma)$ for any given $h\in\spa^>(B)$. Certainly, we also have $h\in\spa^>(A)$, so that we can write (\ref{derint}) by $A$-differentiability. There, we can choose, for $a$ as in the proposition, $\alpha(z) = \Delta_{a;z}^A f(\Sigma)$ and its complex conjugate for $\b\beta(\b{z})$. Contour deformation from $\p A^-$ to $\p B^-$ can be performed since the singularity at $a$ is never crossed. Using Corollary \ref{corolint}, we find $B$-differentiability and $\alpha$ is in the holomorphic $B$-class. In particular, $\alpha(z)$ is the unique member identified with $\Delta_{a,z}^B f(\Sigma)$.
\eproof

A simple corollary of Proposition \ref{propoAB} is the following statement.
\begin{corol}\label{corolAB}
If a function $f$ is both $A$-differentiable and $B$-differentiable at $\Sigma$ for some simply connected domains $A$ and $B$ whose complements have non-empty intersection, $\hC\setminus(A\cup B)\neq\emptyset$, then
\beq
	\Delta_{a;z}^B f(\Sigma) = \Delta_{a;z}^A f(\Sigma)
\eeq
for any $a\in\hC\setminus (A\cup B)$.
\end{corol}
\proof By Proposition \ref{propoAB}, we know that $f$ is $C$-differentiable for any simply connected $C$ that includes $A\cup B$. Then, from Proposition \ref{propoAB} again, $\Delta_{a;z}^A f(\Sigma) = \Delta_{a;z}^C f(\Sigma)$ and $\Delta_{a;z}^B f(\Sigma) = \Delta_{a;z}^C f(\Sigma)$. \eproof

This corollary is very close to Proposition \ref{propoindep} proved above, but does not directly imply it and is not directly implied by it. Proposition \ref{propoindep} tells us about the equality of certain directional derivatives (hence of the conformal derivatives on a subspace) for any simply connected domains $A$ and $B$ with non-empty intersection; whereas Corollary \ref{corolAB} tells us about the equivalence of the holomorphic derivatives (but the corresponding conformal derivatives may act on very different spaces), with the requirement that the exteriors of $A$ and $B$ have non-empty intersection.

\begin{figure}
\bc
\includegraphics[width=6cm,height=3cm]{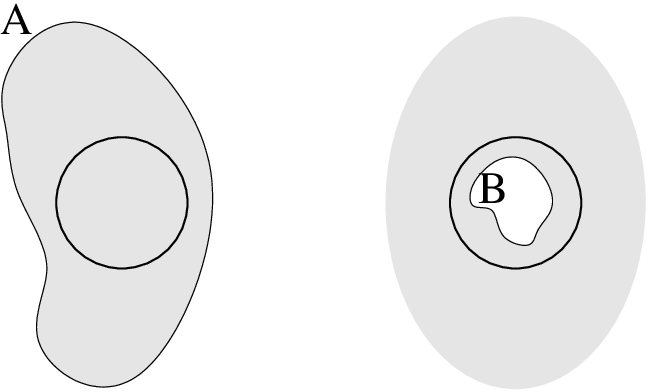}
\ec
\caption{An example: $\Sigma$ is the unit circle centered at 0, $\Omega$ is the space of smooth loops in $\hC$. There are two natural sectors for the derivative of any differentiable function $f$ at $\Sigma$: $[A]$ and $[B]$. The fundamental holomorphy region of the sector $[A]$ is $\hC\setminus\cl{\uD}$, and that of the sector $[B]$ is $\uD$.}
\label{figsect}
\end{figure}

Let $\au f(\Sigma)$ be the set of all simply connected domains $A$ such that $f$ is $A$-differentiable at $\Sigma$. Define an equivalence relation $\simeq$ between elements of $\au f(\Sigma)$ by requiring that $A\simeq B$ if they are such that $\hC\setminus A$ and $\hC\setminus B$ have non-empty intersection, completing by transitivity. Then, from the relation $\simeq$ defined in Subsection \ref{ssectcontdual} and from Corollary \ref{corolAB}, we immediately find:
\begin{corol}\label{corolABgen}
Let $A,B\in \au f(\Sigma)$, and $a\in \hC\setminus A$, $b\in \hC\setminus B$. If $A\simeq B$, then $\Delta_{a,\cdot}^A f(\Sigma) \simeq \Delta_{b,\cdot}^B f(\Sigma)$.
\end{corol}
We can partition the set $\au f(\Sigma)$ into equivalence classes $\au_i f(\Sigma)$ (parametrized by an index $i$) under $\simeq$, which we will call {\em sectors}. When there is no ambiguity, we will denote by $[A]$ the sector $\au_i f(\Sigma)$ such that $A\in \au_i f(\Sigma)$. If there is more than one sector in the partition, we will say that the derivative of $f$ at $\Sigma$ is {\em multi-partite}; otherwise, we will say that it is {\em complete}. See figure \ref{figsect} for an example.

For any sector $\au_i f(\Sigma)$, we can define the corresponding {\em fundamental set} $\cap_{A\in \au_i f(\Sigma)} A$. The complement of this set in $\hC$ is a region of holomorphy of the holomorphic derivative $\Delta_{a;z}^A f(\Sigma)$ for any $A\in\au_i f(\Sigma)$ (up to, possibly, a pole of order 3 at $z=a$), and will be called the {\em fundamental holomorphy region} of the sector. Note that the fundamental set contains the non-trivial singularity structure of the holomorphic derivatives, and that this singularity structure is a characteristic of the sector.

\subsection{Global stationarity and global holomorphic derivatives}

The most important concept for the applications that we will be looking at is that of {\em global holomorphic derivative}, or simply global derivative: it is the holomorphic derivative in the cases where $f$ is invariant under displacements of $\Sigma$ by global conformal transformations in a neighborhood of the identity. The holomorphic derivative $\Delta_{a;z}^A f(\Sigma)$ then has no singularity at $a$, and it does not depend on $a$ or on the particular choice of the domain in the sector $[A]$. As a consequence, the global derivative also enjoys simple transformation properties.

We will say that $f$ is {\em globally stationary} at $\Sigma$ if it is stationary at $\Sigma$ along any one-parameter subgroup of global conformal maps. We have:
\begin{theorem}\label{theoglobal}
If $f$ is $A$-differentiable at $\Sigma$ for some simply connected domain $A$ and globally stationary at $\Sigma$, then the quadratic differential (as a function of $z$)
\beq
	\Delta^{[A]}_z f(\Sigma) := \Delta^A_{a;z} f(\Sigma)
\eeq
is independent of the choice of domain in the sector $[A]$, and is independent of $a\in\hC\setminus A$, for any $A$. Also, it is a non-singular quadratic differential on the fundamental holomorphy region associated to that sector. For any given $A$, the unique member of the holomorphic $A$-class that is non-singular on $\hC\setminus A$ is $\Delta^{[A]}_z f(\Sigma)$.
\end{theorem}
\proof If $f$ is globally stationary at $\Sigma$, then $\nabla^A f(\Sigma)\in \spa^{>*}_\perp(A)$. Hence, from (\ref{dualglob}), Corollary \ref{corolABgen}, and the fact $\simeq$ is the usual analytic continuation for non-singular quadratic differentials, the theorem follows.
\eproof

The quadratic differential in this theorem is the {\em global holomorphic derivative} of $f$ at $\Sigma$ associated to the sector $[A]$. From (\ref{transmob}), we immediately obtain:
\begin{theorem}\label{theotransglob}
Let $A$, $A'$, $f$, $f'$, $\Sigma$ and $\Sigma'$ be as in Proposition \ref{propog}, with $g=G:A\tto A'$ a M\"obius map. If $f'$ is $A'$-differentiable at $\Sigma'$ and globally stationary at $\Sigma'$, then
\beq \label{transglob}
	\Delta_z^{[A]} f (\Sigma) = (\p G(z))^2 \Delta_{G(z)}^{[A']} f'(\Sigma').
\eeq
\end{theorem}
\proof The assumptions of the theorem imply that $f$ is globaly stationary at $\Sigma$, and from Proposition \ref{propog}, is $A$-differentiable at $\Sigma$. Hence, the global derivative of $f$ at $\Sigma$ exists by Theorem \ref{theoglobal}. Equation (\ref{transmob}) gives the result.
\eproof

Since $\Delta^{[A]}_z f(\Sigma)$ is a quadratic differential, the content of the relation (\ref{transglob}) is that the global derivative of $f'\circ G$ at $\Sigma$ can be obtained from that of $f'$ at $G(\Sigma)$ by conformally transporting it by $G$, if $G$ is a M\"obius map. This is completely analogous to the behavior of the operator $\p_z$ (the derivative at a point), where a coefficient appears under a change of coordinates ($z$ is a coordinate). The coordinate-independent way to describe it is to discuss the operator $h(z) \p_z$ where $h$ is a $(-1,0)$-differential. In our case, a change of coordinates around $\Sigma$ is reproduced by a conformal transport of the quadratic differential (note that we do not have to choose coordinates on the Riemann sphere itself). Hence, the global holomorphic derivative is a derivation on functions on the $A^*$-manifold in a neighborhood of $\Sigma$, which is not valued in quadratic differentials, but rather in ``moving'' quadratic differentials, fixed once a coordinate system at $\Sigma$ is chosen. A coordinate-independent description could be one where coordinate systems around $\Sigma$ and on the Riemann sphere are tied together - but this provides an invariant definition under M\"obius maps only. We will see below in what situation this can be made into a truly invariant definition.

Note that if in fact $f$ is invariant under global conformal transformations (not just stationary), then we can also use $f'=f$ in (\ref{transglob}). This relation then has another interpretation: it indicates how to transport the quadratic differential under M\"obius transport of the points at which we differentiate.

Using global derivatives, we can obviously write
\beq
	\nabla_h f(\Sigma) = \int_{z:\vec\p A^-}\dd z\, h(z) \Delta_z^{[A]} f(\Sigma) +
		\int_{z:\vec\p A^-} \bd\b{z} \,\b{h}(\b{z}) \b\Delta_{\b{z}}^{[A]} f(\Sigma)
\eeq
for any $h$ holomorphic on $A$. Deforming the contours, the analytic properties of the global derivative make it possible to relate it directly to the directional derivative in the direction given by the vector field (in global coordinates)
\beq\label{hw}
	h^{(w)}(z) = \frc1{w-z}
\eeq
for $w\in\hC\setminus A,\; w\neq\infty$. Indeed, we have
\beq\label{hwd}
	\nabla_{h^{(w)}} f(\Sigma) = \Delta_w^{[A]} f(\Sigma) + \b\Delta_{\b{w}}^{[A]} f(\Sigma)
\eeq
and the inverse equation can be written in different ways, for instance:
\beq\label{ds}
	\Delta_w^{[A]} f(\Sigma) = \frc12 \sum_{\pm} e^{\mp i\pi/4} \,\nabla_{e^{\pm i\pi/4}h^{(w)}} f(\Sigma)
		= \frc1{2\pi} \int_0^{2\pi} d\theta\, e^{-i\theta}\, \nabla_{e^{i\theta}h^{(w)}} f(\Sigma).
\eeq

This in fact suggests that we should define a regularized holomorphic derivative in general, even if there is no global stationarity, as follows, in global coordinates on the Riemann sphere:
\beq\label{eqdefiholom}
	\Delta^{[A]}_z f(\Sigma) := \lt(\Delta^A_{z;w} f(\Sigma) - \mbox{\rm singular terms about $w=z$}\rt)_{w=z}
	= \Delta^A_{\infty;z} f(\Sigma) \quad \mbox{(if $\infty \not\in A$)}.
\eeq
The equality follows from Corollary \ref{corolABgen}: we simply have to evaluate, for $w\in \C$ in the fundamental holomorphy region associated to $[A]$, the contour integral $\int_{z:\vec\p A^-}\dd z\, \frc1{w-z} \Delta_{a;z}^{A} f(\Sigma)$ in two ways: putting $a=\infty$, or putting $a=w$. In both cases, the contour can be deformed to a small contour surrounding $w$, giving the result stated. What we obtain is a function on the Riemann sphere (here in global coordinates) that is not naturally a quadratic differential; further it is not in general an element of an holomorphic $A$-class. However, it indeed specializes to the global derivative (in global coordinates) when there is global stationarity. Naturally, Equations (\ref{hwd}), (\ref{ds}) hold as well when there is no global stationarity, using the regularized holomorphic derivative.

We now introduce an object associated to the global derivatives that will turn out to play an important role below. We know, by Proposition \ref{propog}, that under the conditions of that proposition, $(\p g(z))^2 \Delta_{g(z)}^{[g(A)]} (f\circ g^{-1})(g\cdot \Sigma) = u(z) + \Delta_z^{[A]} f(\Sigma)$ for some $u\in \spa^<(A)$. That is, there is an object $u$ that tells us how to transport the global derivative in order to reproduce a change of coordinates around $\Sigma$ under conformal maps that are not M\"obius. We refer to this object as the {\em $A$-connection of $f$ at $\Sigma$ associated to a conformal transformation $g:A\to B$}, and denote it by:
\beq\label{deficon}
	\conn_{z;g}^{[A]} f(\Sigma) := \Delta_z^{[A]} f(\Sigma) - (\p g(z))^2 \Delta_{g(z)}^{[g(A)]} (f\circ g^{-1})(g\cdot \Sigma).
\eeq
This, as function of the point $z$ on the Riemann sphere, defines a quadratic differential on $A$. It tells us how to transform the global holomorphic derivative upon change of coordinates at $\Sigma$ that are not M\"obius maps -- it is not simply a change-of-coordinate transformation of a quadratic differential, but involves a supplementary term, a quadratic differential on the complement domain.

Theorem \ref{theotransglob} is simply saying that the $A$-connection is zero for M\"obius maps. Using the analytic properties of the global derivative, the $A$-connection can be written in an integral form:
\beq\label{intcon}
	\conn_{w;g}^{[A]} f(\Sigma) = \int_{z:\vec\p A^-} \frc{dz}{w-z}
		(\p g(z))^2 \Delta_{g(z)}^{[g(A)]} (f\circ g^{-1})(g\cdot \Sigma)
	\quad (w\in A).
\eeq
A similar integral form holds for the global derivative:
\beq
	\Delta_{w}^{[A]} f(\Sigma) = \int_{z:\vec\p A^-} \frc{dz}{w-z} (\p g(z))^2 \Delta_{g(z)}^{[g(A)]} (f\circ g^{-1})(g\cdot \Sigma)
	\quad (w\in\hC\setminus A).
\eeq
From the definition of the $A$-connection, it is easy to derive its transformation property:
\beq\label{transcon}
	\conn_{w;g_1\circ g_2}^{[A]} f(\Sigma) =
		\conn_{w;g_2}^{[A]} f(\Sigma) + (\p g_2(w))^2\, \conn_{g_2(w);g_1}^{[g_2(A)]}(f\circ g_2^{-1})(g_2\cdot \Sigma).
\eeq
By the duality $A\leftrightarrow \hC\setminus A$ in (\ref{dualglob}), it seems natural to interpret the $A$-connection in terms of conformal $\hC\setminus A$-derivatives. Theorem \ref{theotransgen} below, which is our main theorem for this section, indeed gives the $A$-connection such an interpretation.

If we wanted to generalize (\ref{transglob}) to any transformation $g$ that is conformal on $A$, we would obviously encounter problems in establishing the analytic structure on $\hC\setminus A$, since there $g$ is not analytically constrained. In order to resolve this, we rather attempt to generalize it to transformations that are conformal on $\hC\setminus A$, i.e. {\em outside} $A$. We cannot directly use the class transformation properties that we have introduced, because they hold for transformations conformal on $A$. In effect, though, what we will use are similar transformation properties, but for derivatives associated to {\em doubly-connected} domains (although we do not explicitly introduce all the details of this kind of derivative). This is ultimately the reason, in the theorem below, for asking for certain continuity properties of the derivatives: such continuity properties would guarantee the existence of the doubly-connected-domain derivative.
\begin{theorem}\label{theotransgen}
Consider two simply connected domains $A$ and $B$ such that $\hC\setminus A \subset B$ (see, e.g. figure \ref{figsect}). Consider a conformal map $g:B\tto B'$ connecting a $A\cap B$-neighborhood of $\Sigma$ to a $A'\cap B'$-neighborhood of $\Sigma' = g\cdot \Sigma$, with $A' = \hC\setminus g(\hC\setminus A)$. Consider a function $f$ on the $A\cap B$-neighborhood of $\Sigma$, and define $f' = f\circ g^{-1}$. Suppose that:
\begin{enumerate}
\item $f'$ is both $A'$-differentiable and $B'$-differentiable at $\Sigma'$ and globally stationary at $\Sigma'$;
\item all directional $A'$-derivatives (resp. $B'$-derivatives) exist uniformly on a $B'$-neighborhood (resp. $A'$-neighborhood) of $\Sigma'$;
\item all directional $A'$-derivatives (resp. $B'$-derivatives) are $B'$-continuous (resp. $A'$-continuous) at $\Sigma'$;
\end{enumerate}
(in both points 2 and 3, one of the two possibilities only needs to be assumed). Then $f$ is $A$-differentiable at $\Sigma$, and for $w\in \hC\setminus A$,
\beq \label{transgen}
	\Delta_w^{[A]} f (\Sigma) - (\p g(w))^2 \,\Delta_{g(w)}^{[A']} f'(\Sigma') =
		\conn_{w;g}^{[B]} f(\Sigma).
\eeq
\end{theorem}
\proof For simplicity, we consider only the case where neither $A$ nor $A'\cap B'$ contain $\infty$, and where $w\neq\infty$. This is without loss of generality: it can always be achieved by applying a global conformal transformation on the domains and by conjugating $g$ by such a transformation. We also use global coordinates. Let us consider the limit
\[
	\lim_{\eta\to0}
			\frc{f (g_{\eta}\cdot \Sigma) - f(\Sigma)}{\eta} =	
	\lim_{\eta\to0}
			\frc{f'((g\circ g_{\eta}\circ g^{-1})\cdot\Sigma') - f'(\Sigma')}{\eta}
\]
where $(g_\eta:\eta>0)\in \spag(A)$, which we can write as $g_\eta = \id + \eta h_\eta$ with $h_\eta \to h\in \spa(A)$ compactly on $A$. Writing $g_{\eta}' = g\circ g_{\eta} \circ g^{-1} = \id + \eta h_{\eta}'$, we have that, for all $\eta$ small enough, 1) $g_{\eta}'$ is conformal on $A' \cap B_\eta'$ with $B_\eta'\to B'$ as $\eta\to0$, 2) $h_{\eta}'$ is holomorphic on $A'\cap B_\eta'$, and 3) $h_\eta'$ compactly tends to $h' = (\p g\, h)\circ g^{-1}$ as $\eta\to0$. The theorem of appendix \ref{appfact} shows that we can write $g_{\eta}' = g_{\eta;A}' \circ g_{\eta;B}'$, where $g_{\eta;B}'$ is conformal on $B_\eta'$ and $g_{\eta;A}'$ is conformal on $\hC\setminus g_{\eta;B}'(\hC\setminus A')$. It also shows that we have, for $z\in A' \cap B_\eta'$,
\beqa
	g_{\eta;B}'(z) &=& z + \eta\int_{y:\vec\p (B_\eta')^-} \dd y\,\frc{\p g_{\eta;B}'(y) h_{\eta}'(y)}{
			g_{\eta;B}'(y) - g_{\eta;B}'(z)} \n
	g_{\eta;A}'(z) &=& z + \eta\int_{y:\vec\p (A')^-} \dd y\, \frc{\p g_{\eta;B}'(y) h_{\eta}'(y)}{
			g_{\eta;B}'(y) - z}. \no
\eeqa
Then, $g_{\eta;B}'$ tends to $\id$ as $\eta\to0$ (in the $B'$-topology). Hence, we find that $(g_{\eta;B}':\eta>0)\in \spag(B')$ and $(g_{\eta;A}':\eta>0)\in\spag(A')$, with
\beqa
	\p (g_{\eta;B}':\eta>0) &=& \int_{y:\vec\p (B')^-} \dd y\,\frc{h'(y)}{y - z} =: h_A'(z) \n
	\p (g_{\eta;A}':\eta>0) &=& \int_{y:\vec\p (A')^-} \dd y\,\frc{h'(y)}{y - z} =: h_B'(z). \no
\eeqa
Note that $h_A'(z) + h_B'(z) = h'(z)$ for $z\in A'\cap B'$, and that $h_A'\in\spa(A')$ and $h_B'\in\spa(B')$. Then, we have
\beqa
	\lim_{\eta\to0}\frc{f'(g_\eta'\cdot \Sigma') - f'(\Sigma')}{\eta} &=&
		\lim_{\eta\to0}\frc{f'(g_{\eta;A}'\cdot g_{\eta;B}'\cdot \Sigma') - f'(g_{\eta;B}'\cdot \Sigma')}{\eta}
		+ \lim_{\eta\to0}\frc{f'(g_{\eta;B}'\cdot \Sigma') - f'(\Sigma')}{\eta} \n
	&=& \nabla_{h_A'}f'(\Sigma') + \nabla_{h_B'}f'(\Sigma') \label{reshahb}
\eeqa
where we used uniformity of the existence of the limit $\lim_{\eta\to0} \frc{f'(g_{\eta;A}'\cdot \t\Sigma) - f'(\t\Sigma)}{\eta}$ for $\t\Sigma$ in a $B'$-neighborhood of $\Sigma'$, as well as $B'$-continuity of the resulting directional derivative $\nabla_{h_A'}f'(\t\Sigma)$. Clearly, we could as well have written $g_{\eta}' = g_{\eta;B}' \circ g_{\eta;A}'$, where $g_{\eta;A}'$ is conformal on $A'$ and $g_{\eta;B}'$ is conformal on $\hC\setminus g_{\eta;A}'(\hC\setminus B_\eta')$. Repeating the process by essentially interchanging $A$ and $B$, we would obtain again the equation above, except that it would be under the conditions of the uniform existence of the limit $\lim_{\eta\to0} \frc{f'(g_{\eta;B}'\cdot \t\Sigma) - f'(\t\Sigma)}{\eta}$ for $\t\Sigma$ in a $A'$-neighborhood of $\Sigma'$, as well as $A'$-continuity of the resulting directional derivative $\nabla_{h_B'}f'(\t\Sigma)$. Since both $h_A'$ and $h_B'$ are continuous linear functionals of $h'$, we have shown $A$-differentiability of $f$ at $\Sigma$.

Then, with (\ref{hw}) and $w\in \hC\setminus A$, we have, using (\ref{ds}),
\[
	\Delta_w^{[A]} f(\Sigma) 
		= \frc12 \sum_{s=\pm} e^{-is\pi/4} \lim_{\eta\to0}
			\frc{f (g_{\eta}\cdot \Sigma) - f(\Sigma)}{\eta}
\]
where
\[
	g_{\eta}(z) = z + \eta e^{i s\pi/4} h^{(w)}(z).
\]
Here, for lightness of notation, we keep the dependence on $w$ and $s$ implicit. Using the general result (\ref{reshahb}), with $h'$ expressed in terms of $h^{(w)}$ instead of $h_\eta$, this gives
\beqa
	\Delta_w^{[A]} f(\Sigma) &=& \int_{z:\vec\p (A')^-} \dd z\, h_A'(z)\, \Delta_z^{[A']} f'(\Sigma') +
		\int_{z:\vec\p (B')^-} \dd z\, h_B'(z)\, \Delta_z^{[B']} f'(\Sigma') \n
	&=& \int_{z:\vec\p (A')^-} \dd z\, h'(z)\, \Delta_z^{[A']} f'(\Sigma') +
		\int_{z:\vec\p (B')^-} \dd z\, h'(z)\, \Delta_z^{[B']} f'(\Sigma') \n
	&=& \int_{z:\vec\p A^-} \dd z\, (\p g(z))^2 h^{(w)}(z) \,\Delta_{g(z)}^{[A']} f'(\Sigma') +
		\int_{z:\vec\p B^-} \dd z\, (\p g(z))^2 h^{(w)}(z) \,\Delta_{g(z)}^{[B']} f'(\Sigma') \n
	&=& (\p g(w))^2 \,\Delta_{g(w)}^{[A']} f'(\Sigma') +
		\int_{z:\vec\p B^-} \dd z\, (\p g(z))^2 h^{(w)}(z) \,\Delta_{g(z)}^{[B']} f'(\Sigma'). \no
\eeqa
In the second step we used holomorphy of $h_A'$ on $A'$ and of $h_B'$ on $B'$, as well as the respective holomorphy on complement domains of the global derivatives $\Delta_z^{A'} f'(\Sigma')$ and $\Delta_z^{B'} f'(\Sigma')$ along with the behavior $O(z^{-4})$ as $z\to\infty$ (we only need $O(z^{-1})$). In the last step, we evaluated the first integral similarly using holomorphy. The theorem follows from (\ref{intcon}).
\eproof

Hence, the theorem gives us the somewhat surprising relation
\beq
	\lt(\Delta_w^{[A]} - \Delta_w^{[B]}\rt) f (\Sigma)
	= (\p g(w))^2 \,\lt(\Delta_{g(w)}^{[\hC\setminus g(\hC\setminus A)]} -
	\Delta_{g(w)}^{[g(B)]} \rt) (f\circ g^{-1})(g\cdot \Sigma).
\eeq
This is surprising, because the quadratic differentials involved have very different analyticity properties, and the derivatives involved are with respect to very differente families of conformal maps. An immediate and useful consequence of the theorem is the following corollary.
\begin{corol} \label{coroltransgeninv} In the context of theorem \ref{theotransgen}, if the $B'$-derivative of the function $f'$ at $\Sigma'$ is zero, or equivalently if the $B$-derivative of $f$ at $\Sigma$ is zero (that is, if $f'$ is $B'$-stationary at $\Sigma'$, or equivalently if $f$ is $B$-stationary at $\Sigma$), then
\beq\label{transgeninv}
	\Delta_w^{[A]} f (\Sigma) = (\p g(w))^2 \,\Delta_{g(w)}^{[A']} f'(\Sigma').
\eeq
\end{corol}
That is, the quadratic-differential transformation property holds exactly in this case. Since $A$ and $B$ have no exterior point in common, they can be in different sectors (as in the situations that we will be considering), in which case this corollary is a somewhat non-trivial result (if $A$ and $B$ are in the same sector, then the corollary is trivial because both sides vanish).

Formula (\ref{transgeninv}) means that when there is {\em domain stationarity}, there is an invariant definition of the conformal derivative whereby changes of coordinates on the $A^*$-local manifold that are conformal on neighborhoods of $\hC\setminus A$, are tied with changes of coordinates of the quadratic differential on the Riemann sphere. If there is invariance under transformations conformal on $B$, instead of merely domain stationarity, then $f'=f$ in (\ref{transgeninv}). Hence, in this case we can interpret the formula as a transport formula under such changes of coordinates: the quadratic differential is transported in the natural way.

Applications of formulas (\ref{transgen}) and (\ref{transgeninv}) to CFT and CLE indeed involve an interpretation as transport equations, instead of change-of-coordinate equations. Assuming that $f$ is also $A'$-differentiable and globally stationary at $\Sigma'$, Equation (\ref{transgen}) can be written
\beqa\label{leq}
	\chrg_{w;g}^{[A]} f(\Sigma) &:=& \Delta_w^{[A]} f (\Sigma) - (\p g(w))^2 \,\Delta_{g(w)}^{[A']} f(\Sigma') \\ &=&
		\conn_{w;g}^{[B]} f(\Sigma) - (\p g(w))^2 \Delta_{g(w)}^{[A']} (f-f\circ g^{-1})(\Sigma')\no
\eeqa
If there is no domain stationarity (let alone conformal invariance), this object still has a nice interpretation in CFT. From the definition of $\chrg_{w;g}^{[A]} f(\Sigma)$ (\ref{leq}), it transforms as
\[
	\chrg_{w;g_1\circ g_2}^{[A]} f(\Sigma) = \chrg_{w;g_2}^{[A]} f(\Sigma) +
		(\p g_2(w))^2 \chrg_{g_2(w);g_1}^{[\hC\setminus g_2(\hC\setminus A)]} f(g_2\cdot\Sigma).
\]
Moreover, $\chrg_{w;g}^{[A]} f(\Sigma)$ is holomorphic on the fundamental holomorphy region of the sector $[A]$, and it vanishes if $g$ is a global conformal map and there is global invariance. If $\chrg_{w;g}^{[A]}f(\Sigma)$ is in fact independent of $\Sigma$, then the analytic structure, transformation properties and vanishing for global conformal maps can be solved by the Schwarzian derivative $\{g,w\}$,
\beq\label{chrg}
	\chrg_{w;g}^{[A]}f(\Sigma) = \frc{c}{12} \{g,w\}.
\eeq
It turns out that this form is explicitly observed in the example of the stress-energy tensor in the next section (see Subsection \ref{ssectonept}), as well as in the example of the CLE construction in \cite{I}. In these cases, $c$ corresponds to the central charge of the model.

\subsection{Other simple relations}

Most of the usual properties of derivatives of course hold for conformal derivatives. For instance, we have the chain rule for the holomorphic derivative $\Delta_{a;z}^A$: with a differentiable function $F:\R\to \R$,
\beq\label{chainrule}
	\Delta_{a;z}^{[A]} (F\circ f)(\Sigma) = F'(f(\Sigma)) \Delta_{a;z}^{[A]} f(\Sigma).
\eeq
Moreover, it is also possible to study functions of many arguments: $\Sigma = \Sigma_1 \times \Sigma_2$, for instance. As usual, if 1) both partial derivatives of $f$ with respect to $\Sigma_1$ and $\Sigma_2$ exist, 2) all partial directional derivatives with respect to $\Sigma_1$ exist uniformly in a neighborhood of $\Sigma_2$, and 3) all partial directional derivatives with respect to $\Sigma_1$ are continuous at $\Sigma_2$, then we have that $f$ is differentiable as a function of $\Sigma$, and that
\beq\label{multiarg}
	\Delta_{a;z\,|\,\Sigma}^{[A]} f(\Sigma) =
		\Delta_{a;z\,|\,\Sigma_1}^{[A]} f(\Sigma_1\times \Sigma_2) + \Delta_{a;z\,|\,\Sigma_2}^{[A]} f(\Sigma_1\times \Sigma_2).
\eeq
Here, we introduced the notation $|\,\Sigma$ in order to indicate the argument with respect to which the derivative is taken. Finally, the application to functions valued in a general real-linear space is obtained by linearity. There is the usual subtlety when taking complex-valued functions $f:\Omega\to \C$, as they can be seen as valued in the two-dimensional real-linear space $\R^2\cong\C$, or in the one-dimensional complex-linear space $\C$. Since the conformal derivative itself is a real-linear operator, this does not lead to any ambiguity. But the holomorphic derivative extends the field by mapping real-valued functions to complex-valued functions, hence can more naturally be seen as a linear operator on the complex-linear space of complex-valued functions. That is, in the natural definition
\beq
	\Delta_{a;z}^{[A]} f(\Sigma) = \Delta_{a;z}^{[A]} ({\rm Re}\circ f)(\Sigma) +
		i\Delta_{a;z}^{[A]} ({\rm Im}\circ f)(\Sigma),
\eeq
we may see the imaginary number $i$ as an element of the field, not simply a basis element for the linear space $\R^2$. The natural definition for the anti-holomorphic derivative simply takes the complex conjugate of the real and imaginary parts separately:
\beq
	\b\Delta_{\b{a};\b{z}}^{[A]} f(\Sigma) = \b\Delta_{\b{a};\b{z}}^{[A]} ({\rm Re}\circ f)(\Sigma) +
		i\b\Delta_{\b{a};\b{z}}^{[A]} ({\rm Im}\circ f)(\Sigma).
\eeq

\sect{Applications to CFT} \label{sectWard}

\subsection{Singularity structure and conformal Ward identities}

Lie-group invariance in field theory often implies the existence of local fields satisfying local conservations laws. Conformal invariance in two dimensions, in particular, leads to the existence of the stress-energy tensor, whose conservation laws essentially imply that it must be composed of two components: one holomorphic and one anti-holomorphic \cite{BPZ,F84} (for tutorials, see, for instance, \cite{Gins,DFMS97}). In quantum field theory, conservation laws are broken at the locations of other local fields, in a way that is exactly determined by their transformation properties -- this is encoded into the {\em Ward identities}. Accordingly, conformal Ward identities express the fact that the stress-energy tensor, in conformal field theory, is not holomorphic/anti-holomorphic at the location of other local fields: there are poles, whose coefficients are fixed by the conformal transformation properties of these local fields \cite{BPZ,F84}.

In general, the transformation properties of local fields can be written as
\beq\label{trfields}
	(g\cdot\Or)(g(z)) = \sum_i q_{i}(\p g(z), \p^2 g(z), \ldots, \p^n g(z))\Or^{(i)}(g(z)),
\eeq
where $q_i(x_1,x_2,\ldots,x_n)$ are of the form $x_1^{\alpha_i} \b{x}_1^{\beta_i}$ times polynomials in $x_1,\b{x}_1,x_2,\b{x}_2,\ldots,x_n,\b{x}_n$, and the sum over $i$ is finite. This has the meaning that if the model is considered on a domain $C$ or on the Riemann sphere $C=\hC$, then correlation functions are invariant,
\beq\label{invCFT}
	\bra \prod_{j=1}^n (g\cdot\Or_j)(g(z_j))\ket_{g(C)} = \bra \prod_{j=1}^n \Or_j(z_j)\ket_C,
\eeq
for transformations $g$ conformal on $C$ (we use global coordinates and take the positions of the fields to be different from $\infty$ for simplicity). It is important that, by locality, the properties (\ref{trfields}) do not depend on the region $C$ where the theory is considered, or on the boundary conditions. Note that we obtain constraints on the correlation functions by taking $g(C)=C$; otherwise (\ref{invCFT}) can be seen as defining correlation functions on other domains of $\hC$ (or on more general open sets if $g$ is multiply-valued on $C$), once they are known on some standard domain (say $C=\uH$).

Let us denote by $T(w)$ and $\b{T}(\b{w})$, respectively, the fields representing the holomorphic and anti-holomorphic components of the stress-energy tensor at the point $w$. In order to extract the pole structure, one may use the formal relation\footnote{This relation may be made precise by understanding it as holding inside appropriate correlation functions, or more algebraically as a relation in the context of vertex operator algebras.}
\beq\label{sing}
	(g_\eta\cdot \Or)(g_\eta(z)) =
		\lt(1 + \eta\,\oint_z \lt[ \dd w\,h(w)\,T(w) + \bd \b{w}\,\b{h}(\b{w}) \b{T}(\b{w})\rt] + o(\eta)\rt) \Or(z)
\eeq
expressing the fact that the contour integral of the stress-energy tensor generates infinitesimal conformal transformations. Here, $(g_\eta:\eta>0)\in \spag(A)$ for some domain $A$ such that $z\in A$, and $h = \p(g_\eta:\eta>0)$. In particular, if $q(\p g(w),\p^2g(w),\ldots) = (\p g(w))^\delta (\b{\p}\b{g}(\b{w}))^{\t{\delta}}$ (this is the transformation property of primary fields of conformal dimensions $\delta$ and $\t{\delta}$), one immediately finds the pole structures
\[
	T(w) \Or(z) \sim \frc{\delta}{(w-z)^2}\Or(z) + \frc1{w-z} \frc{\p}{\p z} \Or(z),\quad
	\b{T}(\b{w}) \Or(z) \sim \frc{\t\delta}{(\b{w}-\b{z})^2}\Or(z) + \frc1{\b{w}-\b{z}} \frc{\p}{\p \b{z}} \Or(z).
\]
Relation (\ref{sing}) uniquely fixes the pole structure of $T(w) \Or(z)$ (and its conjugate) at $w=z$ for any transformation properties (\ref{trfields}).

Note that the stress-energy tensor itself transforms in a determined way \cite{BPZ}:
\beq\label{transTCFT}
	(g\cdot T)(g(w)) = (\p g(w))^2 T(g(w)) + \frc{c}{12} \{g,w\}
\eeq
(and similarly for the anti-holomorphic component) where $\{g,w\}$ is the Schwarzian derivative:
\beq
	\{g,w\} = \frc{\p^3 g(w)}{\p g(w)} - \frc32 \lt(\frc{\p^2g(w)}{\p g(w)}\rt)^2.
\eeq
The constant $c$ is a characteristic of the CFT model under study (it is the central charge of the Virasoro algebra satisfied by the modes of the stress-energy tensor).

\subsection{Boundary conditions and extended conformal Ward identities}

If one considers a CFT model on the Riemann sphere $\hC$, then it is possible to express fully and exactly the effect of inserting the stress-energy tensor into a correlation function: the exact function is deduced from the exact pole structure, along with holomorphy away from the poles on the whole Riemann sphere \cite{BPZ}. For instance, if $\Or_j$ are primary fields of conformal dimensions $\delta_j,\t{\delta}_j$, then
\beqa
	\bra T(w) \prod_{j=1}^n \Or_j(z_j) \ket_\hC &=&
	\sum_{j=1}^n \lt(\frc{\delta_j}{(w-z_j)^2} + \frc1{w-z_j} \frc{\p}{\p z_j} \rt) \bra \prod_{j=1}^n \Or_j(z_j) \ket_\hC \n
	\bra \b{T}(\b{w}) \prod_{j=1}^n \Or_j(z_j) \ket_\hC &=&
	\sum_{j=1}^n \lt(\frc{\t\delta_j}{(\b{w}-\b{z}_j)^2} + \frc1{\b{w}-\b{z}_j} \frc{\p}{\p \b{z}_j} \rt)
		\bra \prod_{j=1}^n \Or_j(z_j) \ket_\hC. \no
\eeqa
In order to actually fix the overall constant (allowed by holomorphy), one uses the fact that correlation functions factorize at large distances (here we use the Euclidean distance), and that the average of the stress-energy tensor on the plane $\C$ is 0 by rotation covariance.

If one considers a CFT model on domains in $\hC$, however, there is no immediate simple formula, because the analytic structure of the stress-energy tensor outside the domain is not fixed; rather, certain boundary conditions are fixed. Yet, on simply connected domains it is still possible to obtain simple formulae, where the effect of the boundary conditions is obtained by putting local fields outside of the domain of definition (the resulting formulae only depend on the CFT model through the central charge, something that is true only for simply connected domains). Indeed, in general, if the real line is a boundary component, then the boundary condition along it was found by Cardy \cite{C84} to be simply $T(x)=\b{T}(x),\,x\in\R$. Hence, for a CFT model on the upper half-plane $\uH$, we may analytically extend correlation functions $\bra T(w) \prod_{j=1}^n \Or_j(z_j) \ket_\uH$, as functions of $w$, towards the lower half-plane $\uL$, and fix the pole structure there -- this is a type of {\em reflection} property. The pole structure on $\uL$ is simply given by the known pole structure of $\bra \b{T}(\b{w}) \prod_{j=1}^n \Or_j(z_j) \ket_\uH$ found for $w\in\uH$, but with the variable $\b{w}$ replaced by $w$. For instance, with primary fields we have
\[
	\bra T(w) \prod_{j=1}^n \Or_j(z_j) \ket_\uH =
	\sum_{j=1}^n \lt(\frc{\delta_j}{(w-z_j)^2} + \frc1{w-z_j} \frc{\p}{\p z_j} 
	+ \frc{\t\delta_j}{(w-\b{z}_j)^2} + \frc1{w-\b{z}_j} \frc{\p}{\p \b{z}_j}\rt) \bra \prod_{j=1}^n \Or_j(z_j) \ket_\uH.
\]
Here again we used the fact that correlation functions factorize at large distances, and that the average of the stress-energy tensor on $\uH$ is 0 by covariance. Then, we can simply apply a conformal transformation mapping $\uH$ to any other simply connected domain, and use the transformation property (\ref{transTCFT}).

Since the transformation property (\ref{transTCFT}) involves the Schwarzian derivative, in general the insertion of the stress-energy tensor for models on simply connected domains $C$ will involve a ``disconnected term'', equal to $\bra T(w)\ket_C \bra \prod_{j=1}^n \Or_j(z_j) \ket_C$. It is convenient to consider {\em connected correlation functions},
\beq
	\bra T(w) \prod_{j=1}^n \Or_j(z_j) \ket_C^{(c)} = \bra T(w) \prod_{j=1}^n \Or_j(z_j) \ket_C
	 - \bra T(w)\ket_C \bra \prod_{j=1}^n \Or_j(z_j) \ket_C.
\eeq
Connected correlation functions transform as if the holomorphic component of the stress-energy tensor were a primary field of conformal dimensions $2,0$. That is, we have
\beqa\label{invconnected}
	(\p g(w))^2 \bra T(g(w)) \prod_{j=1}^n (g\cdot\Or_j)(g(z_j))\ket_{g(C)}^{(c)}
		&=& \bra T(w) \prod_{j=1}^n \Or_j(z_j)\ket_C^{(c)} \\
	(\b\p \b{g}(\b{w}))^2 \bra \b{T}(\b{g}(\b{w})) \prod_{j=1}^n (g\cdot\Or_j)(g(z_j))\ket_{g(C)}^{(c)}
		&=& \bra \b{T}(\b{w}) \prod_{j=1}^n \Or_j(z_j)\ket_C^{(c)}. \no
\eeqa
Note that, in particular, we find
\[
	\bra T(w) \prod_{j=1}^n \Or_j(z_j) \ket_\uH^{(c)} = \bra T(w) \prod_{j=1}^n \Or_j(z_j) \ket_\uH,\quad
	\bra T(w) \prod_{j=1}^n \Or_j(z_j) \ket_\hC^{(c)} = \bra T(w) \prod_{j=1}^n \Or_j(z_j) \ket_\hC.
\]

For models defined on multiply-connected domains $C$, there is no simple way of extracting the exact stress-energy tensor insertions. This ultimately is due to the fact that the exact form, in the multiply-connected case, depends on the boundary conditions on the various boundary components. However, let us consider $C$ to be a disk with circular holes inside it -- this can always be achieved by conformal transformations. In this case, it is possible to reduce the effect of the boundary conditions to single isolated singularities in each of the components of the complement $\hC\setminus C$. Indeed, it is always possible to map conformally the disk or the complement of any of its holes to $\uH$. Applying the boundary condition $T(x) = \b{T}(x),\,x\in\R$, by reflection we can extend the region where the analytic structure is known beyond $\uH$ -- only poles will appear. Mapping back to $C$, we have extended the region towards the exterior of the disk or the inside of the holes. Since in order to map disks (or global transform thereof) to $\uH$ we can use global conformal transformations, there is no Schwarzian derivative involved, and no additional singularity is incurred through the transformation properties of the local fields. Repeating the process, we can extend the region up to single points (where poles accumulate) in each component of $\hC\setminus C$. At these points, additional singularities may be present. These additional singularities contain all the information about the boundary conditions on each boundary component. For instance, for the one-point function $\bra T(w)\ket_C$, we find analyticity everywhere except for such single isolated singularities in each component of $\hC\setminus C$. For connected correlation functions, however, we expect there to be no additional singularities: connected correlation function can be evaluated exactly simply by adding the poles coming from the local fields and all their reflective images (this is expected to form a convergent series).

The exact determination of connected correlation functions of the stress-energy tensor, in terms of correlation functions not involving it, is what we will refer to as the {\em extended conformal Ward identities}. In a sense, they not only tell us about the singularities produced by local fields, but also about those associated to the domain boundary.

\subsection{Extended conformal Ward identities from conformal derivatives}

We do not yet have all the tools to assess the multiply-connected case, but we may show how the extended conformal Ward identities are expressed using conformal derivatives in the case where the region of definition $C$ is $\hC$ or a simply connected domain thereof.

In order to apply conformal differentiability on connected correlation functions, we need to specify the space $\Omega$ on which the correlation functions are seen to act. Let us fix a positive integer $n$ representing the fixed number of local fields in the correlation functions. Local fields, in our context, are naturally seen as forming a linear space ${\cal F}$ over some ring of functions on $\C$; in this sense, then, the transformation properties (\ref{trfields}) make any conformal transformation $g$ into an endomorphism of ${\cal F}$. Since these transformation properties only involve finitely many coefficients, it is sufficient to assume that ${\cal F}$ is finite-dimensional. Denote by ${\cal D}$ the space of simply connected domains $\cup \{\hC\}$, that is, the regions of definition that we look at. We consider $2n+1$-tuplets
\beq\label{Sigma}
	\Sigma = (C;z_1,\ldots,z_n;\Or_1,\ldots,\Or_n)\in {\cal D}\times \C^n\times {\cal F}^{\otimes n}
\eeq
and take $\Omega$ to be the subspace determined by restricting $z_j\in C$ and $z_i\neq z_j$ for $i\neq j$. Then, clearly we define the function $f:\Omega\to \C$ via
\beq\label{fSigma}
	f(\Sigma) = \bra \prod_{j=1}^n \Or_j(z_j) \ket_{C}.
\eeq

The family of conformal transformations acting on $\Omega$ that we consider is that of all maps conformal on $C$, as well as all maps in a $A$-neighborhood of the identity for any simply connected domain $A\supset (\hC\setminus C) \cup S$ for $S=\{z_1,z_2,\ldots,z_n\}\subset C$. The former set of maps acts on $C$ in the natural way, $g\cdot C= g(C)$, and the latter set may also be seen as acting on $C$: we define $g\cdot C$ in this case to be $\hC$ if $C=\hC$, and to be the simply connected domain bounded by $g(\p C)$ and containing $\{g(z_j):j=1,\ldots,n\}$ otherwise (if $g$ is near enough to the identity, it is single valued on $\p C$). Then, we define the action of $g$ on $\Omega$, for $g$ as above, via
\beq\label{actSigma}
	g\cdot \Sigma = (g\cdot C;g(z_1),\ldots,g(z_n);g\cdot\Or_1,\ldots,g\cdot\Or_n).
\eeq
Note that this indeed gives an action consistent with the composition of conformal maps.

We have:
\begin{theorem} \label{theomain} With $\Sigma$, $C$ and $S$ as above, and with $\hC_w = \hC\setminus \cl{N(w)}$ where $N(w)$ is a simply connected open neighborhood of $w$ in $C\setminus S$ (see figure \ref{figinsert}):
\begin{itemize}
\item[A.] The $\hC_w$-global holomorphic derivative of $f$ at $\Sigma$ exists.
\item[B.] Connected correlation functions of the stress-energy tensor components on $C$ can be expressed as global holomorphic derivatives of $f$ at $\Sigma$:
\beq
	\bra T(w) \prod_{j=1}^n \Or_j(z_j) \ket^{(c)}_C = \Delta_w^{[\hC_w]} f(\Sigma),\quad
	\bra \b{T}(\b{w}) \prod_{j=1}^n \Or_j(z_j) \ket^{(c)}_C = \b\Delta_{\b{w}}^{[\hC_w]} f(\Sigma).
\eeq
\end{itemize}
\begin{figure}
\bc
\includegraphics[width=2.5cm,height=3cm]{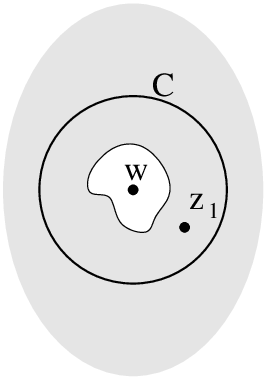}
\ec
\caption{The domain $C$ (bounded by the bold circle and containing $w,z_1$), and the domain $\hC_w$ (shaded area).}
\label{figinsert}
\end{figure}

\end{theorem}
\proof The initial observation is that, from CFT, correlation functions $\bra \prod_{j=1}^n \Or_j(z_j)\ket_\uH$ and $\bra \prod_{j=1}^n \Or_j(z_j)\ket_\hC$ are infinitely differentiable functions of $\{z_1,z_2,\ldots,z_n\}\in \R^{2n}$ (for non-colliding points $z_j$ lying in the domain of definition $\uH$ or $\hC$). Along with (\ref{invCFT}), this implies that $f$ is $\hC_w$-differentiable. Note that by (\ref{invCFT}), we may in fact reduce the space of ``inequivalent'' correlation functions (i.e.\ that are not related by a product of functions of the individual positions) to a finite number of copies of open sets in $\uH$ and $\C$ (the {\em moduli space}), so that $\hC_w$-differentiability is essentially reduced to differentiability on a finite-dimensional manifold. Moreover, from (\ref{invCFT}) we clearly have
\beq\label{invf}
	f(g\cdot \Sigma) = f(\Sigma) \quad\forall\quad \mbox{$g$ conformal on $C$}.
\eeq
In particular, we have global stationarity, hence by Theorem \ref{theoglobal} the global derivative exists. This proves A.

For the proof of B, first note that if $C$ contains $\infty$, then $\Delta_w^{[\hC_w]} f(\Sigma)$ vanishes as $w\to\infty$, so that we have the correct asymptotic condition. The proof then involves three steps: showing that $\Delta_w^{[\hC_w]} f(\Sigma)$ and $\b\Delta_{\b{w}}^{[\hC_w]} f(\Sigma)$ transform in agreement with (\ref{invconnected}), showing that they have the correct analytic structure for $w\in C$, and showing that for $C=\uH$, they satisfy the correct boundary condition on $\R$. Then, by the discussion above, the equalities follow.

It will sometimes be convenient to consider the real and imaginary parts of $f$ separately; we will denote by ${\bf f}$ the vector formed by these separated functions: ${\bf f} = ({\rm Re}\circ f,{\rm Im}\circ f)$.

The first step follows immediately from Theorem \ref{theotransglob} in the case where $C = \hC$. Otherwise, it uses Corollary \ref{coroltransgeninv} (which follows from Theorem \ref{theotransgen}) as follows. The invariance formula (\ref{invf}) implies that for any $g:B\tto B'$ conformal on a domain $B\supset \cl{C}$, we have that $(f\circ g)(\Sigma)=f(\Sigma)$, this being true on a $B$-neighborhood of $\Sigma$, and also that $f\circ g$ is $B$-differentiable at $\Sigma$. Moreover, the conditions of Theorem \ref{theotransgen}, with $A=\hC_w$ and $B$ as said, are clearly satisfied. These considerations in fact hold true for the real and imaginary parts of $f$ independently (i.e.\ hold for ${\bf f}$). Replacing $\hC\setminus g(\cl{N(w)})$ by $\hC_{g(w)}$ (which we can of course do), we have
\beq\label{eip}
	\Delta_w^{[\hC_w]} {\bf f} (\Sigma) = (\p g(w))^2 \,\Delta_{g(w)}^{[\hC_{g(w)}]} {\bf f}(g\cdot \Sigma).
\eeq
For simplicity, let us restrict to $C = \uD$. Let us write $g(z) = \t{g}(rz)$ for some $r<1$, and consider $\t{g}$ conformal on $C$. The right-hand side of (\ref{eip}) exists at $r=1$ and is continuous as $r\to 1^-$ for any fixed $w\in C$ and $\Sigma\in\Omega$. This is because the $\hC_{g(w)}$-neighborhood and $\hC_{g(w)}$-derivative of ${\bf f}$ at $g\cdot \Sigma$ exist for all $r\in (0,1]$, because $g\cdot \Sigma$ represents, as a function of $r$, a continuous path lying entirely in the moduli space for $r\in(0,1]$, and because we have infinite differentiability on the moduli space (and recall that the moduli space is a manifold). Hence we may take the limit $r\to 1^-$ on both sides. Similar arguments may be provided for other choices of $C$, and we find that (\ref{eip}) holds for all $g$ conformal on $C$. Hence, it holds for $f$ itself, in agreement with (\ref{invconnected}), which concludes the first step.

For the second step, we write $\Sigma = \Sigma_0\times \Sigma_1\times \Sigma_2 \times \cdots \times \Sigma_n$, with $\Sigma_0=C$ and $\Sigma_j = (z_j,\Or_j)$. The $\hC_w$-differentiability and continuity conditions leading to (\ref{multiarg}) certainly hold at $\Sigma$, hence we have, for instance in the holomorphic case,
\[
	\Delta_w^{[\hC_w]} f(\Sigma) = \sum_{j=0}^n \Delta_{a;w\,|\,\Sigma_j}^{\hC_w} f(\Sigma).
\]
Here, we may take $a$ to be any fixed point in $N(w)$. On the right-hand side, every term may have a pole of order up to 3 at $w=a$ (if $\infty\in \hC_w$), but they cancel out since on the left-hand side there is no such singularity. Hence, we may simply omit these singularities. Here, it is convenient to simply subtract these singularities in each term on the right-hand side, hence to use the regularized holomorphic derivatives. That is, we have
\beq\label{partial}
	\Delta_w^{[\hC_w]} f(\Sigma) = \sum_{j=0}^n \Delta_{w\,|\,\Sigma_j}^{\hC_w} f(\Sigma).
\eeq
For the first term, involving $\Sigma_0$, note that we can extend the space of conformal maps acting on $C$ simply by omitting the requirement that they be conformal on $S$. Then, we see that we have $A$-differentiability as function of $\Sigma_0$ for any simply connected $A$ such that $\p C\in A$, so that the first term provides a holomorphic contribution to $\Delta_w^{\hC_w}f(\Sigma)$ for $w\in C$. If $C = \hC$, then obviously no $\Sigma_0$-derivative needs to be taken, so the first term is 0. Hence, the singularities in $C$ may only come from the derivatives with respect to $\Sigma_j$ for $j=1,\ldots,n$. The conformal derivative formula (\ref{derint}) can be written, in the case of the first factor $\Sigma_1$ for instance, as
\beqa\label{eqs1}
	\lefteqn{\bra g_\eta \cdot \Or_1(g_\eta(z_1)) \prod_{j=2}^n \Or_j(z_j) \ket_{C}} && \\ &=&
		f(\Sigma) +
		\eta \int_{z:\vec\p \hC_w} \dd z\,h(z)\,
			\Delta_{a;z\,|\,\Sigma_1}^{[\hC_w]} f(\Sigma) +
		\eta \int_{z:\vec\p \hC_w} \bd\b{z}\,\b{h}(\b{z})\,
			\b\Delta_{\b{a};\b{z}\,|\,\Sigma_1}^{[\hC_w]} f(\Sigma)
		+ o(\eta) \no
\eeqa
for any $\{g_\eta:\eta>0\}\in\spag(\hC_w)$. From (\ref{ds}) and the form of the coefficient functions $q_i$ in (\ref{trfields}), it is clear that only finite-order poles can occur, and, in $\hC$, only at $w=z_1$, in the function $\Delta_{a;z\,|\,\Sigma_1}^{[\hC_w]} f(\Sigma)$ (except for the singularity at $z=a$). In order to establish exactly what these poles are, we only have to compare (\ref{eqs1}) with (\ref{sing}). Since in (\ref{eqs1}) we may evaluate the contour integrals by deforming them in $\hC_w$, we see that the singularity of $\Delta_{a;z\,|\,\Sigma_1}^{[\hC_w]} f(\Sigma)$ in $\hC_w$ is uniquely fixed by that of $T(z)\Or_1(z_1)$, and we conclude that it is the correct singularity for the stress-energy tensor. The singularity at the point $w=a$, outside $\hC_w$, is taken away in $\Delta_{w\,|\,\Sigma_1}^{[\hC_w]} f(\Sigma)$, without affecting other singularities. Since similar statements hold for anti-holomorphic counterparts, we find that $\Delta_{w}^{[\hC_w]} f(\Sigma)$ and $\b\Delta_{\b{w}}^{[\hC_w]} f(\Sigma)$ have the correct pole structure in $C$. This concludes the second step.

For the third step, let us specialize to $C=\uH$. We will show that
\beq\label{toshow}
	\Delta_w^{[\hC_w]} {\bf f}(\Sigma) =
		\sum_{j=1}^n \lt(\Delta_{w\,|\,\Sigma_j}^{[\hC_w]} + \b\Delta_{w\,|\,\Sigma_j}^{[\hC_w]}\rt) {\bf f}(\Sigma).
\eeq
Since the right-hand side is holomorphic on $\hC\setminus (S\cup \b{S})$, the left-hand side may be analytically extended to that region, and we may specialize to $w\in \R$. There, by complex conjugation, we see that $\Delta_w^{\hC_w} {\bf f}(\Sigma) = \b\Delta_w^{\hC_w} {\bf f}(\Sigma)$, hence putting together real and imaginary parts we obtain the correct boundary condition on $\R$.

In order to show (\ref{toshow}), let us consider derivatives with respect to $\Sigma_0$, and write, using (\ref{ds}),
\[
	\Delta_{w\,|\,\Sigma_0}^{[\hC_w]} {\bf f}(\Sigma) = \int_0^{2\pi} \frc{d\theta\,e^{-i\theta}}{2\pi}
		\nabla_{h_{w,\theta}\,|\,\Sigma_0} {\bf f}(\Sigma)
\]
where
\[
	h_{w,\theta} = \frc{e^{i\theta}}{w-z}\quad (\theta\in\R).
\]
Consider $g_\eta(z) = z + \eta h_{w,\theta}(z)$. We can find ${\cal G}'=\{g_\eta':\eta>0\}\in\spag(\hC_{\b{w}})$ such that $g_\eta(\R) = g_\eta'(\R)\;\forall\;\eta>0$. Indeed, for any $g_\eta$ there is a unique $g_\eta'$ conformal on $\uH$ such that $g_\eta'(\R) = g_\eta(\R)$, with, for instance, the normalization $g_\eta'(z) \sim z + O(1/z)$ as $z\to\infty$. Consider $G_\eta:=g_\eta^{-1}\circ g_\eta'$, which is such that $G_\eta(\R)=\R$. For any fixed $z$ away from $w$ and $\b{w}$, $g_\eta'(z)$ and $G_\eta(z)$ have convergent Taylor expansions in $\eta$ about $\eta=0$. It is easy to see that with $G_\eta(z) = z - \eta e^{i\theta}/(w-z) - \eta e^{-i\theta}/(\b{w}-z) + O(\eta^2)$ we find $g_\eta'(z) = z - \eta e^{-i\theta}/(\b{w}-z) + O(\eta^2)$. Hence, $\p{\cal G}' = -h_{\b{w},-\theta}$, so that we have
\[
	\nabla_{h_{w,\theta}\,|\,\Sigma_0} {\bf f}(\Sigma) = -\nabla_{h_{\b{w},-\theta}\,|\,\Sigma_0} {\bf f}(\Sigma).
\]
We then get
\[
	\Delta_{w\,|\,\Sigma_0}^{[\hC_w]} {\bf f}(\Sigma) = -
		\int_0^{2\pi} \frc{d\theta\,e^{i\theta}}{2\pi}
		\nabla_{h_{\b{w},\theta}\,|\,\Sigma_0} {\bf f}(\Sigma) = -\b\Delta_{w\,|\,\Sigma_0}^{[\hC_{\b{w}}]} {\bf f}(\Sigma).
\]
But since $\b\Delta_{w}^{[\hC_{\b{w}}]} {\bf f}(\Sigma) = 0$ by (\ref{invf}), we obtain
\[
	\Delta_{w\,|\,\Sigma_0}^{[\hC_w]} {\bf f}(\Sigma) = \sum_{j=1}^n \b\Delta_{w\,|\,\Sigma_j}^{\hC_w} {\bf f}(\Sigma),
\]
which, along with (\ref{partial}), shows (\ref{toshow}). \eproof

The proof of Theorem \ref{theomain} makes it clear that we can subdivide the action of the global derivative into its action on the various arguments of the correlation functions. In particular, if $D_n(w)$ is the differential operator representing the pole structure of the holomorphic component of the stress-energy tensor,
\[
	\bra T(w) \prod_{j=1}^n \Or_j(z_j)\ket^{(c)}_C \sim D_n(w) \bra \prod_{j=1}^n \Or_j(z_j)\ket_C,
\]
with on $\hC$
\[
	\bra T(w) \prod_{j=1}^n \Or_j(z_j)\ket^{(c)}_\hC = D_n(w) \bra \prod_{j=1}^n \Or_j(z_j)\ket_\hC,
\]
then we have on simply connected domains $C$
\[
	\bra T(w) \prod_{j=1}^n \Or_j(z_j)\ket^{(c)}_C = \lt(D_n(w) + \Delta_{w\,|\,C}^{\hC_w}\rt) \bra \prod_{j=1}^n \Or_j(z_j)\ket_C,
\]
where the regularized holomorphic derivative acts on $C$ in the way explained above (i.e. it acts on $\p C$ by conformal transformations). For instance, with primary fields, and re-writing the regularized holomorphic derivative as an integral, we have
\beq\label{otherform}
	\bra T(w) \prod_{j=1}^n \Or_j(z_j) \ket^{(c)}_C =
	\lt(\sum_{j=1}^n\lt(\frc{\delta_j}{(w-z_j)^2} + \frc1{w-z_j} \frc{\p}{\p z_j}\rt)
	+ \int_{z:-\vec\p C^-} \dd z\, \frc1{w-z} \Delta_{z\,|\,C}^{\hC_w}\rt) \bra \prod_{j=1}^n \Or_j(z_j) \ket_C.
\eeq
In this form, it is apparent that the boundary of the domain of definition can be considered as a ``continuum of zero-dimensional primary fields'', where $\Delta_{z\,|\,C}^{\hC_w}$ can be interpreted as the holomorphic derivative with respect to the part of the boundary near $z$.

\subsection{One-point average}\label{ssectonept}

It is well known \cite{F84} that the one-point average of the stress-energy tensor on a domain $C$ can be expressed via a variation of the partition function $Z_C$ on $C$ under a metric change, in a neighborhood of the flat, Euclidean metric (see (\ref{basicform})). This seems to point to an expression of the one-point average in terms of a conformal derivative. It turns out that the global holomorphic derivative $\Delta_w^{\hC_w}$ used to reproduce the extended conformal Ward identities above can be used to reproduce as well the one-point average. However, the one-point average cannot simply be the global derivative of a partition function: the latter is not globally stationary in general. There is a particular ratio of partition functions, which we call {\em relative partition function}, that is globally stationary (in fact, globally invariant).  This paticular ratio is inspired by results in the context of CLE \cite{I}, where a relative partition function is defined using CLE renormalized probability functions.

\begin{figure}
\bc
\includegraphics[width=2.5cm,height=3cm]{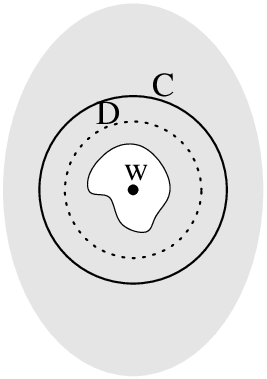}
\ec
\caption{The domains $C$ and $D$ (bounded by the bold circles and with $w\in D\subset C$), and the domain $\hC_w$ (shaded area), in formula (\ref{onept}).}
\label{figrelat}
\end{figure}

The relative partition function $Z(C|D)$, depending on two domains $C$ and $D$ with $\cl{D}\subset C$, is defined as
\beq
	Z(C|D) = \frc{Z_C Z_{\hC\setminus\cl{D}}}{Z_{C\setminus\cl{D}}}
\eeq
(up to a constant factor). Our main formula in this subsection is that the one-point average can be expressed as
\beq\label{onept}
	 \bra T(w)\ket_C = \Delta_{w\,|\,\p C\cup \p D}^{[\hC_w]} \log Z(C|D)
\eeq
for $w\in D$ (see figure \ref{figrelat}).

The derivative is taken with respect to $\p C\cup \p D$, where the action of conformal maps in a $\hC_w$-neighborhood of $\id$ is by conformal transformation of the set $\p C\cup \p D$ (the transformed set can then be interpreted as boundaries of two new simply connected domains $C'$ and $D'$ with $\cl{D'}\subset C'$). In particular, the result of the derivative is independent of the domain $D$. This means that, in general, correlation functions can be expressed as
\[
	\bra T(w) \prod_{j=1}^n \Or_j(z_j) \ket_C =
		Z(C|D)^{-1} \Delta_{w\,|\,\Sigma\times \p D}^{[\hC_w]} \Big( Z(C|D) f(\Sigma)\Big)
\]
where the derivative is with respect to $\Sigma$ (\ref{Sigma}) (which includes $C$, with an action on $\p C$ in agreement with that above) and $\p D$, and conformal maps act on $\Sigma\times \p D$ as $g\cdot (\Sigma\times \p D) = g\cdot \Sigma \times g(\p D)$, with $g\cdot\Sigma$ as in (\ref{actSigma}). By the transformation property (\ref{transTCFT}), we see that the global derivative
in (\ref{onept}) transforms in agreement with (\ref{leq}) and (\ref{chrg}), where $c$ is the central charge. The derivation of (\ref{onept}) is reported in appendix \ref{apponept}; it is based on CFT arguments, and is far from being of mathematical rigor\footnote{This derivation appeared already in the preprint \cite{I}.}. A more mathematically rigorous derivation for corresponding objects in the context of CLE is found in \cite{I}.

\sect{Conclusions} \label{sectConc}

In the present paper, we have developed the notion of derivative on groupoids of conformal maps with (almost) a local manifold structure near to the identity. The main conclusion is that some fundamental aspects of CFT appear naturally in this general geometric context. More precisely, our first main result is that such a derivative, when there is global stationarity, can be described using an object with a clean analytic structure and simple transformation properties under conformal maps. Our second main result is that this object is, in fact, intimately related to the stress-energy tensor: it exactly reproduces the extended conformal Ward identities (the conformal Ward identities and the boundary conditions) for connected correlation functions. We also provided arguments indicating that it also reproduces the one-point averages of the stress-energy tensor.

Natural paths for extending and applying this work include: studying the full differentiable manifold of conformal maps (i.e. not just around the identity); extending to higher derivatives and stress-energy tensor descendants (a first part of which was done in \cite{DVOA}); generalizing to manifolds involving Lie groups so as to connect with other holomorphic symmetry currents in CFT; applying the formalism to deduce the form of the stress-energy tensor and other symmetry currents in other probabilistic theories connected to CFT (e.g.~the Gaussian field); analyzing derivatives of functions characterizing other mathematical objects that may have close links with conformal maps; generalizing to a description of massive QFT.

{\bf Acknowledgments}

I would like to thank D. Bernard and J. Cardy for asking questions in August 2008 that led to the present work, as well as C. Hagendorf, K. Kytola and, in particular, D. Meier for encouragements and helpful discussions. I also acknowledge support form an EPSRC First Grant, ``From conformal loop ensembles to conformal field theorey" EP/H051619/1.

\appendix

\sect{Proof of structure of the continuous dual $\spa^*$} \label{appproof}

We give an elementary proof of lemma \ref{lemlf}.

\noindent \proof Certainly, the set $\{\Upsilon H_{n,s}:n=0,1,2,\ldots,\,s=\pm\}$ is part of the characterisation of $\Upsilon$. Consider $h^{(N)} = \sum_{n=0,\ldots,N,\,s=\pm} c_{n,s} H_{n,s}$. Then $\Upsilon h^{(N)}$ is given by (\ref{lflincomb}) by linearity. But since $\lim_{N\to\infty} h^{(N)}= h$ in $\spa$, we have $\lim_{N\to\infty} \Upsilon h^{(N)} = \Upsilon h$ by continuity. This shows (a).

If $\{a_{n,s}:n=0,1,2,\ldots,\,s=\pm\}$ is a series of coefficients for some $h$, $a_{n,s} = c_{n,s}(h)$, then so is any reassignment of signs of the $a_{n,s}$s, because of absolute convergence of Taylor series in the disk of convergence. Hence, convergence of the r.h.s. of (\ref{lflincomb}) implies that (\ref{seriesconv}) is true (with $b_{n,s} = \Upsilon H_{n,s}$) for all $\Upsilon\in\spa'$. Suppose that we have a sequence of nonnegative reals $\{b_{n,s}:n=0,1,2,\ldots,\,s=\pm\}$ such that the following is {\em not} true: $\exists\, C>0,r\in[0,1)\;|\;\forall\,n\ge0,s=\pm:b_{n,s}\leq Cr^n$. That is, suppose that $\forall\,C>0,r\in(0,1):\exists\, n,s\;|\;b_{n,s}> Cr^n$. Let us construct the function $C(r) = 1/(1-r)$, and the function $n(r)$ that gives the smallest nonnegative integer such that $b_{n(r),s}>C(r)r^n$ for some $s$. Since $C(r)\to\infty$ as $r\to1^-$, then the sequence $N$ of strictly increasing integers that $n(r)$ takes as $r\to1^-$ is an infinite sequence. We also consider the sequence $S$ of the doublets $(n,s)$ for all $n\in N$, with the corresponding values of $s$ such that $b_{n(r),s}>C(r)r^n$. Let us construct the sequence with elements $a_{n,s}$ given by $1/b_{n,s}$ for $(n,s)\in S$, and $0$ otherwise. For any given $n$, there is a $r\in(0,1)$ such that $a_{n,s}<C(r)^{-1} r^{-n}<r^{-n}$; moreover, as $n$ increases, this $r$ increases. Then, for any $r_0\in(0,1)$, we have that $a_{n,s}<r_0^{-n}$ for all $n$ large enough. Hence, $a_{n,s} = c_{n,s}(h)$ for some $h\in\spa$, because $\sum_{n,s} a_{n,s} H_{n,s}(z)$ converges for any $|z|<r_0$. On the other hand, the series $\sum_{n,s}a_{n,s}b_{n,s}$ diverges (is infinite) because $a_{n,s}b_{n,s}=1$ for $(n,s)\in S$ and $0$ otherwise, and $S$ is an infinite sequence. This shows that if (\ref{seriesconv}) holds, then $\exists\, C>0,r\in(0,1)\;|\;\forall\,n\ge0,s=\pm:|b_{n,s}|\leq Cr^n$. As a consequence, any $\Upsilon\in\spa'$ gives rise to a function $\gamma$ in (\ref{Cz}) that is holomorphic on $\hC\setminus\uD$. This shows (b).

Since then (\ref{lfint}) gives rise to the correct action of $\Upsilon$ on the basis, and gives rise to a continuous mapping, by (a) it is true that the class ${\cal C}$ completely characterizes any $\Upsilon\in\spa'$. If two functions $w_1$ and $w_2$ are both in ${\cal C}$, then $w_1-w_2$ is holomorphic on $\uD$, and if additionally both are holomorphic in $\hC\setminus \uD$, then $w_1-w_2$ is holomorphic on $\hC$. Since $w_1(\infty)-w_2(\infty)=0$, it must be that $w_1-w_2=0$. Suppose the relation (\ref{lfint}) holds for all $h$, and both for $\alpha = \alpha_1$ and for $\alpha = \alpha_2$. We can always isolate the holomorphic part by taking linear combinations of the cases with $h$ and with $ih$, so that by subtracting, we have $\int_{z:\vec\p\uD^-} dz\,h(z)\,(\alpha_1(z) - \alpha_2(z))=0$ for all $h\in\spa$. Since $\alpha_1-\alpha_2$ is holomorphic on an annulus with $\p\uD$ as part of its boundary, we can write (by Cauchy's integral formula) $\alpha_1-\alpha_2 = w_1+w_2$ where $w_1$ is holomorphic on $\uD$, and $w_2$ is holomorphic on $\hC\setminus \uD$ (i.e.\ in a neighborhood of this closed set). We are left with $\int_{z:\vec\p\uD^-} dz\,h(z)\,w_2(z)=0$. Taking $h(z) = z^n$ for $n=0,1,2,3,\ldots$, we show that all coefficients of the Taylor expansion of $w_2(z)$ about $\infty$ are zero, hence that $w_2=0$. Hence, $\alpha_1-\alpha_2=w_1\in\spa$, so that $\alpha_1$ and $\alpha_2$ are in the same class. A similar argument holds for $\beta$. Thanks to (b), this shows (c), and then immediately implies (e).

Since any sequence $\{\Upsilon H_{n,s}:n=0,1,2,\ldots,s=\pm\}$ with the condition that $\exists\, C>0,r\in(0,1)\;|\ \forall\,n\ge0,s=\pm:|\Upsilon H_{n,s}|\leq Cr^n$ gives rise to a function (\ref{Cz}) holomorphic on $\hC\setminus\uD$, hence to a continuous functional, and since this condition is a consequence of (\ref{seriesconv}), this shows that (\ref{seriesconv}) is sufficient. Since (\ref{seriesconv}) was shown to be necessary above, this completes the proof of (d).\eproof

\sect{Factorisation of conformal maps on annular domains} \label{appfact}

In this appendix, we work out one result that is needed in the proof of Theorem \ref{theotransgen}. In order to make the derivation clearer, we will employ a simpler notation than what is used in that proof. Consider two simply connected domains $A$ and $B$ such that $\hC\setminus A \subset B$; then $A\cap B$ is an annular domain of $\hC$. If a conformal map $g$ on $A\cap B$ is near enough to the identity, then it can be factorized: we can write it as a composition $g_{A'}\circ g_B$ of a map $g_B$ conformal on $B$ and a map $g_{A'}$ conformal on $A' = \hC\setminus g_B(\hC \setminus A)$. We express this result more precisely as follows (we will use the phrase {\em winding annular subdomain} of an annular domain $D$ to designate an annular subdomain $C\subset D$ that separates the boundary components $\p D$).

\noindent {\bf Theorem}\footnote{There is a more general theorem of factorization, not needing the smallness condition of $g$; a proof of this more general theorem using the uniformization theorem for Riemann surfaces can be found in \cite{Dconf}, and earlier proofs in \cite{Hub66,Kuh70}. But the present theorem is sufficient, and its present proof has some content which may be of interest besides the particular problem at hand.}\\ \em
{\bf I.}\ 
Consider two simply connected domains $A$ and $B$ such that $\hC\setminus A \subset B$. For any compact subset $\alpha\subset A\cap B$ that contains some winding annular subdomain of $A\cap B$, there exists a $r>0$ such that any map $g$ conformal on $A\cap B$ satisfying:
\begin{enumerate}
\item ${\rm max}\big(d(g(z),z):z\in \alpha\big)< r$ where $d(\cdot,\cdot)$ is the distance in the round metric on the Riemann sphere,
\item there are open neighborhoods $N_A\subset A\cap B$ of $\p A$ and $N_B\subset A\cap B$ of $\p B$ such that $g(N_A)\cap g(N_B) = \emptyset$,
\end{enumerate}
is {\em factorizable}: there exist a map $g_B$ conformal on $B$ and univalent on $\hC\setminus A$, and a map $g_{A'}$ conformal on $A' = \hC\setminus g_B(\hC\setminus A)$, such that
\beq\label{fact}
	g = g_{A'} \circ g_B
\eeq
on $A\cap B$.

\em

We may always simplify the problem by considering, instead of $g$, the map $g\circ G_2$ for some fixed M\"obius map $G_2$. Hence, we may assume without loss of generality that $\infty\not\in \cl{A\cap B}$ (in fact we may take $B=\uD$). Likewise, we may consider, instead of $g$, the map $G_1\circ g$ for some fixed M\"obius map $G_1$; then we may assume without loss of generality that $\infty\not\in \cl{g(A\cap B)}$. We will assume these two properties. Further, we may replace the round-metric distance $d(\cdot,\cdot)$ by the plane distance $|\cdot-\cdot|$; we will do this in the following.

Then, we can also modify the maps $g_B$ and $g_{A'}$ without changing $g$ by writing $g= g_{A'} \circ G^{-1} \circ G \circ g_B$ with $G$ another M\"obius map. Thanks to this, we can assume without loss generality that 1) if $\infty\in B$, then $g_B(w) = w + O(1/w)$ as $w\to\infty$, and 2) if $\infty\in A$, then $\infty\in A'$ and $g_{A'}(w) = w + O(1/w)$ as $w\to\infty$ (note that $\infty$ is contained in $A$ or $B$, but not both, by our previous assumption).

\em
\noindent {\bf II.}\ By choosing $g$, $g_B$ and $g_{A'}$ as above, the following integral equations hold:
\beqa
	g_B(z) &=& z + \int_{y:\vec\p B^-} \dd y\, \frc{\p g_B(y) \,(g(y)-y)}{g_B(y) - g_B(z)} \quad (z\in B) \label{iet} \\
	g_{A'}(z) &=& z + \int_{y:\vec\p A^-} \dd y \,\frc{\p g_B(y) \,(g(y)-y)}{g_B(y) - z} \quad (z\in A') \label{gAt}.
\eeqa
The right-hand side of (\ref{iet}) should be understood as the analytic continuation of an expression with contour and argument in a subdomain of $B$ where $g_B$ is univalent.
\em

\noindent {\bf Proof}

We may consider $A\cap B$ and $g$ satisfying the assumptions above, and look for $r$ and $\alpha$ such that the point 1 of part I holds.

Let us first prove that with an appropriate choice of $r$, there must be a winding annular subdomain $C$ of $A\cap B$ where $g$ is univalent.

The minimal distance between $\p B$ and $\p A$ is finite and non-zero. Let us choose $r>0$ such that $\alpha\subset A\cap B$ contains the closure of a winding annular subdomain $C'\subset A\cap B$ with the minimal distance between the two components of $\p C'$ being greater than $4r$. Then, with $|g(z)-z|<r$ for $z\in \alpha$, we now show that the map $g$ is univalent on the winding annular subdomain $C$ with $\p C$ at every point a distance $2r$ from $\p C'$. Indeed, suppose it is not univalent there. Then consider $z_1,z_2\in C$ such that $g(z_1) = g(z_2)$ and $z_1\neq z_2$. Consider also a smooth, simple, unwinding curve $\gamma\in C$ from $z_1$ to $z_2$. Then $g(\gamma)$ is a smooth loop a distance less than $r$ away from $C$. The loop $g(\gamma)$ may have double or higher order points; if it does, we look at the pre-image of these points on $\gamma$ and choose new $z_1$ and $z_2$ such that $g(\gamma)$ is a simple loop. Then, there must be parts of the boundary of $g(C')$ on both simply connected components of $\hC\setminus g(\gamma)$. This is because $g^{-1}$ is conformal in a neighborhood of the loop $g(\gamma)$, hence can be analytically continued from there, and is doubly valued in a neighborhood of $g(z_1) = g(z_2)$. Hence the analytic continuation in any simply connected component of $\hC\setminus g(\gamma)$ must give rise to a branch point, which would map to a non-conformal point of $g$. This has to be shielded by the boundary of $g(A\cap B)$, hence also by the boundary of $g(C')$. Since, then, there are parts of the boundary of $g(C')$ on both simply connected components of $\hC\setminus g(\gamma)$, and since the loop is a distance less than $r$ from $C$, hence more than $r$ from $\p C'$, this means that parts of the boundary $\p C'$ are mapped further away than a distance $r$, a contradiction with the condition $|g(z)-z|<r$ for $z\in \alpha$.

Suppose that we find a factorization $g=g_{\t{A}'}\circ g_{\t{B}}$ on $C = \t{A}\cap \t{B}$, where $g$ is univalent, instead of a factorization (\ref{fact}) on $A\cap B$ (with $\t{A}\subset A$ and $\t{B}\subset B$ simply connected domains, and $\t{A}' = \hC\setminus g_{\t{B}}(\hC\setminus \t{A})$). Suppose also that $g_{\t{B}}$ is univalent on $\t{B}$. Clearly, then, $g_{\t{A}'} = g\circ g_{\t{B}}^{-1}$ is univalent on the annular domain $\t{A}'\cap g_{\t{B}}(\t{B})$, hence on $\t{A}'$.

Then, we may extend the factorization to one that is valid on the whole $A\cap B$ by analytic continuation.

Indeed, the definition $g_{A''} = g\circ g_{\t{B}}^{-1}$ agrees with $g_{\t{A}'}$ on $\t{A}'\cap g_{\t{B}}(\t{B})$, and extends analytically (but not necessarily univalently) to $A'' \cap g_{\t{B}}(\t{B})$ with $A'' = \hC\setminus g_{\t{B}}(\hC\setminus A)$, since $\p A''\subset g_{\t{B}}(\t{B})$, $g_{\t{B}}^{-1}(\p A'') = \p A$, and $g$ is conformal on $A\cap \t{B}$. We will use the same symbol $g_{A''}$ for the resulting analytic map extended to all of $A''$. The conformality conditions $\p g(z) = \p g_{A''}(g_{\t{B}}(z)) \p g_{\t{B}}(z) \neq 0$ and $\p g_{\t{B}}(z)\neq 0$ for $z\in A\cap \t{B}$ further guarantee that $g_{A''}$ is in fact conformal on $A''$.

Hence we have a factorization $g = g_{A''}\circ g_{\t{B}}$ on $A\cap \t{B}$. The definition $g_B = g_{A''}^{-1} \circ g$ agrees with $g_{\t{B}}$ on $A\cap \t{B}$, and we may try to extend it analytically (but not necessarily univalently) to $A\cap B$. The two possible obstructions are if $g$ maps $B\setminus \t{B}$ outside of $g_{A''}(A'')$, the domain of $g_{A''}^{-1}$, or if the analytic continuation is multiply-valued because of the (possible) multiple-valuedness of $g_{A''}^{-1}$. But $g_{A''}((\p A'')^-)$ is $g(\p A^-)$, hence the second condition in part I of the theorem guarantees $g(B\setminus \t{B})$ to be in $g_{A''}(A'')$. Moreover, the analytic continuation will be unique, because two topologically different paths in $g_{A''}(A'')$ between two given points must cross $g_{A''}((\p A'')^-)$, and single-valuedness of $g$ on $C$ as well as the second condition of part I of the theorem forbids the image under $g$ of any path in $B\setminus \t{B}$ to cross $g_{A''}((\p A'')^-)$. We will use the same symbol $g_B$ to designate the resulting analytic map extended to all of $B$. The conformality conditions $\p g(z) = \p g_{A''}(g_{B}(z)) \p g_{B}(z) \neq 0$ and $\p g_{A''}(g_B(z))\neq 0$ for $z\in A\cap B$ further guarantee that $g_{B}$ is in fact conformal on $B$.

Finally, $g_B = g_{\t{B}}$ on $\hC\setminus A$ so that $g_B$ is univalent on $\hC\setminus A$ as well, and so that $A'' = A'$. Then, we recover the factorization (\ref{fact}) (renaming $g_{A''} = g_{A'}$) for the full domain $A\cap B$.

Hence, it is sufficient to assume $g$ to be univalent on $A\cap B$. Moreover, by analytic continuation arguments as above, we may assume that both $\p A$ and $\p B$ are smooth, by replacing $A$ and $B$ by appropriate subdomains. Then, we are looking for a factorization (\ref{fact}) for $g_B:B\to B'$ conformal univalent on $B$ and $g_{A'}$ conformal (and hence also univalent) on $\hC\setminus g_B(\hC\setminus A)$. 

Let us write
\beq\label{gzh}
	g(z) = z + h(z).
\eeq
Certainly, $h$ is holomorphic on $A\cap B$. Suppose that for two conformal maps $g_B$ (univalent conformal on $B$) and $g_{A'}$ (univalent conformal on $A'$), the following equations hold:
\beqa
	g_B(z) &=& z + \int_{y:\vec\p B^-} \dd y\, \frc{\p g_B(y) \,h(y)}{g_B(y) - g_B(z)} \quad (z\in B) \label{ie} \\
	g_{A'}(z) &=& z + \int_{y:\vec\p A^-} \dd y \,\frc{\p g_B(y) \,h(y)}{g_B(y) - z} \quad (z\in A') \label{gA}.
\eeqa
Then we have, for $z\in A\cap B$,
\[
	g_{A'}(g_B(z)) = g_B(z) + \int_{y:\vec\p A^-} \dd y \, \frc{\p g_B(y) \,h(y)}{g_B(y) - g_B(z)}.
\]
Replacing the term $g_B(z)$ by its expression (\ref{ie}), we find
\[
	g_{A'}(g_B(z)) = z + \int_{y:\vec\p (A\cap B)^-} \dd y\, \frc{\p g_B(y) \,h(y)}{g_B(y) - g_B(z)}
		= z + h(z) = g(z)
\]
where the second equation is obtained by Cauchy's theorem. Hence, if the integral equation (\ref{ie}) has a univalent conformal solution on $B$, and that the resulting $g_{A'}$ from (\ref{gA}) is conformal on $A'$, we have found a factorization. This factorization has the properties required for part II of the theorem, hence this would also prove part II.

The integral equation for $g_B$ can be written for its inverse $g_B^{-1}$ as follows (with $B' = g_B(B)$):
\beq\label{appie}
	g_B^{-1}(z) = z - \int_{y:\vec\p (B')^-} \dd y \,\frc{h(g_B^{-1}(y))}{y-z} \quad (z\in B').
\eeq
From there, it is obvious that $g_B^{-1}$ is holomorphic on $B'-\{\infty\}$ (with the correct behavior around $z=\infty$), and we only need to check that $\p g_B^{-1}(z)\neq 0$ there and that $g_B^{-1}$ is univalent on $B'$.

We now show that for $h$ ``small enough'' (as in the theorem), there is a solution giving $g_B$ and $g_{A'}$ with the right properties.

The first part of the strategy is essentially to show that the process of solving the integral equation (\ref{appie}) recursively, starting with $g_B^{-1}(z) = z$, converges to a holomorphic function. Let us write
\[
	(g_B^{-1})_n(z) = z + R_n(z)
\]
with $R_0(z) = 0$ and
\beq\label{recurs}
	R_{n+1}(z) = -\int_{y:\vec\p \t{B}} \dd y\, \frc{h(y+R_n(y))}{y-z}
\eeq
for some simply connected domain $\t{B}\subset B$ (different from the $\t{B}$ in the first part of the proof). Clearly, $R_{n+1}(z)$ is holomorphic for $z$ in a neighborhood of $\cl{\t{B}}$, if $y+R_n(y)\in A\cap B$ for $y\in\p\t{B}$. Let us denote by $|R_n|$ the supremum of $|R_n(z)|$ for $z\in \t{B}$. Let us choose $\t{B}$ as well as another simply connected domain $\t{A}\subset A$ in such a way that $\t{A} \cap \t{B}$ is a non-empty winding annular subdomain of $A\cap B$, and that the smallest distance $S$ between $\p \t{B}$ and $\p B$ is the same as the smallest distance between $\p\t{A}$ and $\p A$. Let us also choose a number $a\in(0,S)$, and denote $S-a={\cal R}>0$. Then, if $|R_n|<{\cal R}$ we indeed find $y+R_n(y)\in A\cap B$ for $y\in\p\t{B}$. We will show by induction that for $h$ small enough on $A\cap B$, the condition $|R_m|<{\cal R}$ for all $m\leq n$ implies $|R_{n+1}|<{\cal R}$, which shows that $R_n$ is holomorphic on a neighborhood of $\cl{\t{B}}$ for all $n$.

Let us then assume that $|R_m|<{\cal R}$ for all $m\leq n$, and consider the differences $\delta_n(z) = R_{n+1}(z) - R_n(z)$. They satisfy
\[
	\delta_{n}(z) = -\int_{y:\vec\p \t{B}} \dd y\, \frc{h(y+R_n(y))-h(y+R_{n-1}(y))}{y-z}.
\]
We now bound the integral involved. For a function $j$ holomorphic on $\t{A}\cap \t{B}$, where $|j(z)|$ has a finite supremum denoted by $|j|$, we can always bound the absolute value of the integral $\int_{y:\vec\p \t{B}} \dd y \,j(y)/(y-z)$ by $\ell |j|/d(z)$ for $z \in \t{B} - \t{A}$, where $\ell$ is the length of $\p \t{B}$ and $d(z)$ is the distance from $z$ to $\p B$ (we imagine taking an integration path along $\p B^-$). For $z\in \t{A}\cap \t{B}$, we can move the integration path away from $z$ before bounding the absolute value, and we can always keep it far enough by bringing it through $z$ if necessary and taking the residue at $y=z$. More precisely, take $\gamma\subset \t{A}\cap\t{B}$ to be the curve at all points equidistant to $\p \t{A}$ and $\p \t{B}$. Consider the components $C_+$ and $C_-$ of $\hC\setminus\gamma = C_+\cup C_-$, the first containing the domain $\t{B}-\t{A}$. For $z\in\cl{C_+}$, we could still take the integration path to be $\p \t{B}^-$; for $z\in C_-$, we could take the integration path to be $\p \t{A}^-$. In the first case, the bound is still $\ell |j|/d(z)$; in the second case, it is $\ell' |j|/d'(z) + |j|$ with $\ell'$ the length of $\p\t{A}$ and $d'(z)$ the distance from $z$ to $\p \t{A}$. We can define the function $q(z)$ by absorbing all factors:
\[
	q(z) = \lt\{\ba{cl}
	d(z)/\ell & z\in \cl{C_+} \\
	d'(z)/(\ell'+d'(z)) & z\in C_-\cap \t{A}\cap\t{B}.
	\ea\rt.
\]
Then, we have
\[
	\lt|\int_{y:\vec\p \t{B}} \dd y \,\frc{j(y)}{y-z}\rt| \leq \frc{|j|}{q(z)}.
\]
Note that $q(z)$ is an increasing function for $z\in\t{B}-\t{A}$ going away from $\t{B}\cap \t{A}$, and that it has an infimum on $\t{B}$ that is greater than 0, i.e.
\[
	q:= {\rm inf}(q(z):z\in\t{B}) \geq {\rm min}\lt(\frc{d}{2\ell},\frc{d}{2\ell'+d}\rt).
\]
where $d$ is the smallest distance between $\p\t{A}$ and $\p\t{B}$.

In our case, we have $j(y) = h(y+R_n(y))-h(y+R_{n-1}(y))$. We write this as
\[
	(R_{n}(y) - R_{n-1}(y))\oint \dd x \frc{h(x)}{(x-y-R_n(y))(x-y-R_{n-1}(y)}.
\]
We can take the $x$ contour to be the oriented boundary $\vec\p X$ of the winding annular subdomain $X$ of $A\cap B$ which is such that $\p X$ is at each point a distance $a/2+{\cal R}$ from $\t{A}\cap \t{B}$. Then, for $y\in\t{A}\cap\t{B}$, we can bound the absolute value of the contour integral by $L_X |h|_X/(a/2)^2$ where $L_X$ is the length of $\p X$, and $|h|_X$ is the supremum of $|h(z)|$ on $X$. Hence, we have
\[
	|j(y)| \leq \gamma_X |h|_X\, |\delta_n(y)|
\]
where $\gamma_X = 4L_X/a^2$. Then, we find, for $z\in\t{B}$,
\[
	|\delta_{n}(z)| \leq \frc{\gamma_X |h|_X \,|\delta_{n-1}|}{q(z)}
\]
where $|\delta_n|$ is the supremum of $|\delta_n(z)|$ for $z\in \t{A}\cap \t{B}$. Since $d(z)$ increases as $z\in\t{B}$ goes away from $\t{A}\cap \t{B}$, the number $|\delta_n|$ is also the supremum of $|\delta_n(z)|$ for $z\in \t{B}$. Solving for this supremum (because by assumption, the bound holds for smaller $n$ as well), this gives
\[
	|\delta_n| \leq \lt(\frc{\gamma_X |h|_X}{q}\rt)^n |\delta_0|.
\]
For $|h|_X$ small enough so that
\beq\label{c1}
	\gamma_X |h|_X<q,
\eeq
we can now bound $|R_{n+1}|$:
\[
	|R_{n+1}| \leq \sum_{m=0}^{n} |\delta_m| \leq \sum_{m=0}^{\infty} \lt(\frc{\gamma_X |h|_X}{q}\rt)^m |\delta_0|
		\leq \frc{|\delta_0|}{1-\frc{\gamma_X |h|_X}{q}}
\]
and since $|\delta_0| = |R_1|$, we have, using the previous method and $R_1(z) = \int_{y:\vec\p \t{B}} \dd y\, \frc{h(y)}{y-z}$,
\[
	|\delta_0|\leq \frc{|h|}{q}
\]
where $|h|$ is the supremum of $h(z)$ on $\t{A}\cap \t{B}$. Hence, we find the bound
\[
	|R_{n+1}| \leq \frc{|h|}{q-\gamma_X |h|_X} \leq \frc{|h|_X}{q-\gamma_X|h|_X}.
\]
Then, for
\beq\label{c2}
	\frc{|h|_X}{q-\gamma_X|h|_X} < {\cal R}
\eeq
we indeed find that $|R_{n+1}|<{\cal R}$, which completes the induction. Note that given the domains $A,B,\t{A},\t{B}$ and the number $a$, the quantities $\gamma_X$, $q$ and ${\cal R}$ are fixed, as well as the domain $X$ determining where the supremum of $|h(z)|$ is taken. Condition (\ref{c2}) can be solved for $|h|_X$, giving
\beq\label{c}
	|h|_X < \frc{q}{\gamma_X + {\cal R}^{-1}}.
\eeq
Hence, this condition is stronger than (\ref{c1}), so is sufficient.

Now we can show that with (\ref{c}) (in fact, only (\ref{c1}) is required), $R_n$ converge uniformly as $n\to \infty$ on $\cl{\t{B}}$, implying that there is a holomorphic solution to (\ref{appie}) with $B'$ replaced by $\t{B}$. Indeed, we have that the sequence $\delta_n(z):n=0,1,2,3,\ldots$ converges uniformly and exponentially to 0 for $z\in \cl{\t{B}}$. Hence, the series $R_\infty(z) = \sum_{n=0}^{\infty} \delta_n(z)$ also converges uniformly for $z\in\cl{\t{B}}$ (because the remainder of the $m^{\rm th}$ partial sum satisfies $\lt|\sum_{n=m}^\infty \delta_n(z)\rt| \leq |\delta_0| (\gamma_X |h|_X/q)^m/(1-\gamma |h|_X/q)\to 0$ as $m\to\infty$ uniformly for $z\in\cl{\t{B}}$). Hence, the limit of the sequence of holomorphic functions $R_n:n=0,1,2,3,\ldots$ is a function $R_\infty$ that is holomorphic on $\t{B}$, and bounded on $\cl{\t{B}}$ by
\beq\label{bdinfty}
	|R_\infty|<\frc{|h|_X}{q-\gamma_X|h|_X}<{\cal R}.
\eeq
The limit can be taken on both sides of (\ref{recurs}), and uniform convergence gives the result.

Let us now consider the function
\beq\label{fctgB}
	g_B^{-1}(z) = z + R_\infty(z),
\eeq
which solves (\ref{appie}) (with $B'$ replaced by $\t{B}$). This function is not only holomorphic, but also conformal on $\t{B}$ for all $|h|_X$ small enough (possibly smaller than the bound (\ref{c})). Indeed, we can bound the absolute value of $\p R_\infty(z)$ by bounding
\[
	\lt|\int_{y:\vec\p \t{B}} dy\,\frc{R_{\infty}(y)}{(y-z)^2}\rt|
\]
using similar techniques as those above, and using (\ref{bdinfty}); this guarantees that for $|h|_X$ small enough, $|\p R_\infty(z)|<1$.

Note that $g_B^{-1}$ in (\ref{fctgB}) compactly tends to the identity as $|h|_X\to0$. Hence, for all $|h|_X$ small enough, there is a domain $\t{B}'$ inside $\t{B}$ where $g_B^{-1}(z)$ is univalent conformal, and this domain tends to $\t{B}$ as $|h|_X\to0$. Then, inverting, we have found a solution $g_B$ to (\ref{ie}), where $B$ is replaced by $B_- = g_{B}^{-1}(\t{B}')$. The function $g_B$ is univalent conformal on $B_-$, and by the construction above, we know that $B_- \subset B$. For $|h|_X\to0$, we have that $B_-\to B$. Hence, by taking $|h|_X$ small enough, we can guarantee that $\p A \subset B_-$. Then, we can construct $g_{A'}$ by (\ref{gA}). The function $g_{A'}$ is analytic on $A' = \hC\setminus g_B(\hC\setminus A)$. The domain $A'$ tends to $A$ as $|h|_X\to0$, so that the function $g_{A'}$ converges compactly to the identity on $A$. Hence, for $|h|_X$ small enough, $g_{A'}$ is univalent conformal on a domain $A_-'\subset A'$. Again by choosing $|h|_X$ small enough, we can guarantee that the domain $A_- = \hC\setminus g_B^{-1}(\hC\setminus A_-')$ has its boundary inside $B_-$, i.e. $\p A_-\subset B_-$, since $A_-\to A$ as $|h|_X\to0$ and $\p A\subset B_-$. That is, we have found a factorization (\ref{fact}) on $A_-\cap B_-$, a winding annular subdomain of $A\cap B$.

By the analytic continuation argument already stated above, and using the fact that $g$ is univalent on $A\cap B$ (by our simplifying assumption), we get a factorization on $A\cap B$.

Hence, we have found a factorization on $A\cap B$, with $g_B$ univalent conformal on $B$ and $g_{A'}$ univalent conformal on $A'$. Note that we can always choose $a$ and ${\cal R}$ small enough so that $X$ is close enough to $\t{A}\cap \t{B}$ in order for $X$ to be inside the compact set $\alpha$. With our previous arguments to extend to the non-univalent case, this completes the proof.
\eproof

\sect{Derivation of the one-point average formula} \label{apponept}

First, we need to describe how a conformal transformation of the domain of definition of a partition function is connected to a change of metric.

A conformal transformation of the domain of definition can be seen as a result of two steps: a re-parametrization of the initial domain, which obviously keeps the partition function invariant but changes the metric by an overall space-dependent factor, and a Weyl transformation that brings back the original metric, but under which the partition function transforms \cite{P81}. We use the standard setup where the trace of the bulk stress-energy tensor is zero, hence the metric we use is flat in the bulk (there is no trace anomaly, see for instance \cite{DFMS97}) -- it can be taken as the Euclidean metric. Then, we consider a partition function on $g(A)$ with that metric, and in the first step, we use $A$ as a parameter space for the domain $g(A)$. The metric it gives on $A$ (in the bulk) is obtained by $|dz|^2 \mapsto |dz|^2 |\p g(z)|^2$. In the second step, the Weyl transformation with a factor $e^{-\sigma(x)} = |\p g(z)|^{-2}$ brings the metric back to the Euclidean metric on $A$, and we have a partition function on $A$.

The transformation of the CFT partition function under a Weyl transformation was found by Polyakov in the context of random surfaces \cite{P81}: for $A$ any appropriate domain (say, any domain with piecewise smooth boundary), we have
\beq\label{transfoZ}
	Z_{g(A)} = e^{\frc{c}{48\pi} S_{\cl{A}}(\sigma)} Z_A
\eeq
where $c$ is the CFT central charge and $S_{\cl{A}}(\sigma)$ is the Liouville action of $\sigma$ on $\cl{A}$,
\beq\label{Liouville}
	S_{\cl{A}}(\sigma) =
	\int_{\cl{A}} d^2x\,\sqrt{\eta} \lt(\frc12 \eta^{ab} \p_a \sigma \p_b \sigma + R\sigma + \mu (e^{\sigma}-1)\rt).
\eeq
Here, $\eta^{ab}$ is the metric on $\cl{A}$ (and $\eta$ is its determinant), $R$ is the associated scalar curvature and $\mu$ is some UV-divergent, non-universal (i.e.\ lattice-model-dependent) scale. Our choice for $\eta^{ab}$ is the Kronecker delta $\delta_{ab}$ in the bulk of $A$.

In general, with curved boundaries, the curvature must have a non-zero contribution supported on the boundary. It is important that the integral in the Liouville action (\ref{Liouville}) covers the boundary of $A$ (which is the meaning of the notation $\int_{\cl{A}}$), so that it gets a non-zero contribution from this term. We will not need a precise description of the boundary term of the metric, but only some properties of the resulting contribution to the Liouville action. We will need that the contribution of the boundary $\p A$ to the Liouville action $S_{\cl{A}}(\sigma)$ only depends on the linear curvature along $\p A$ (besides the value of the function $\sigma$ on $\p A$). We will denote this contribution by $S_{\vec{\p} A}(\sigma)$, where $\vec{\p} A$ is the oriented boundary of $A$, counter-clockwise around the interior of $A$.

Clearly, the partition function in general is not invariant under global conformal maps. Hence, we cannot define the global derivative on it. However, it turns out that there is a certain ratio of partition functions, which we call the {\em relative partition function}, that is globally invariant. This particular ratio was inspired by results in the context of CLE \cite{I}. The relative partition function $Z(C|D)$, depending on two domains $C$ and $D$ with $\cl{D}\subset C$, is defined as
\beq
	Z(C|D) = \frc{Z_C Z_{\hC\setminus\cl{D}}}{Z_{C\setminus\cl{D}}}
\eeq
up to a constant factor. Let us consider a map $g$ that is conformal on $\hC\setminus \cl{D}$ and maps it to a domain of $\hC$. Then, there is also a map $g^\sharp$ conformal on $C$ such that $g^\sharp(\p C) = g(\p C)$. Similarly to the case of correlation functions, we see $Z(C|D)$ as a function of $\p C$ and $\p D$, keeping $\p D$ on the component $C$ of $\hC\setminus \p C$. Let us consider the ratio
\beq
	\frc{Z(g^\sharp(C)|g(D))}{Z(C|D)} = \frc{Z_{g^\sharp(C)}}{Z_C} \frc{Z_{g(\hC\setminus\cl{D})}}{Z_{\hC\setminus\cl{D}}}
		\frc{Z_{C\setminus\cl{D}}}{Z_{g(C\setminus\cl{D})}}.
\eeq
We will argue that this ratio is in fact independent of $\p D$, unity for $g$ a global conformal transformation, and, in some sense, universal. We will then provide further CFT arguments to show, from this formula, that the global derivative $\Delta_w^{[\hC_w]} \log Z(C|D)$ reproduces the stress-energy tensor one-point average.

First, using the transformation property (\ref{transfoZ}), we find
\beqa
	\frc{Z(g^\sharp(C)|g(D))}{Z(C|D)} &=&
	\exp \frc{c}{48\pi} \lt[ S_{\cl{C}}(\sigma^\sharp) + S_{\hC\setminus D}(\sigma) - S_{\cl{C}\setminus D}(\sigma)\rt] \n
	&=& \exp \frc{c}{48\pi} \lt[ S_C(\sigma^\sharp) + S_{\hC\setminus \cl{C}}(\sigma)
		+ S_{\vec\p C}(\sigma^\sharp)-S_{\vec\p C}(\sigma)\rt]
	\label{alsp}
\eeqa
Note the careful inclusion/exclusion of domain boundaries in the Liouville actions. The last expression clearly is independent of $\p D$. Also, suppose $g$ is a global conformal transformation. Then we can choose $g^\sharp = g$ so that $\sigma^\sharp =\sigma$, and we are left with $\exp \frc{c}{48\pi} S_\hC(\sigma)$ (there is no boundary contribution). This is independent of $C$; that it should be 1 can then be obtained simply by sending $C\to\hC$ and $D \to \emptyset$ (assuming continuity). In order to argue that the right-hand side of (\ref{alsp}) is universal in some way, we need to argue that it is mostly independent of $\mu$ (the parameter in the Liouville action (\ref{Liouville})). Since $e^{\sigma} = |\p g|^2$, the $\mu$-terms in $S_C(\sigma^\sharp) + S_{\hC\setminus \cl{C}}(\sigma)$ can be combined into an integration over $\hC$ by change of coordinates; this then provides an overall factor that is independent of $\sigma$. This factor is seen to be 1 by setting $\sigma=0$ (that is, $g=\id$). As for the expression $S_{\p C}(\sigma^\sharp)-S_{\p C}(\sigma)$, there is a non-trivial metric on $\p C$, which we did not specify; but we expect that the resulting combination of $\mu$-terms is universal.

Second, we want to evaluate the derivative $\Delta_{w\,|\,\p C\cup \p D}^{[\hC_w]}$ of $\log Z(C|D)$ and show that it is the stress-energy tensor. Since this is the first derivative, the terms that are quadratic in $\sigma$ in the Liouville actions do not contribute. Also, as we argued above the bulk $\mu$-terms cancel out, and the bulk curvature terms are zero since the bulk metric is flat\footnote{There is a subtlety with the point at $\infty$ when the domain contains it: it takes all the curvature of the Riemann sphere. However, a careful calculation with the metric $d^2x/(1+|z|^2/R^2)^2$, where the curvature is re-distributed, shows that the limit $R\to\infty$ of the curvature term of the Liouville action gives zero contribution to the first derivative.}. This means that we are left only with the boundary contributions to the Liouville actions. Hence we find:
\beq\label{ZZ}
	\Delta_{w\,|\,\p C\cup \p D}^{[\hC_w]} \log Z(C|D) = \frc{c}{48\pi} \Delta_{w\,|\,\sigma}^{[\hC_w]}
		\lt[S_{\vec\p C}(\sigma^\sharp) - S_{\vec\p C}(\sigma)\rt]_{\sigma=0}.
\eeq

Note that with an appropriate renormalization of the partition function $Z_C^R$, we could guarantee that $S_{\hC\setminus \cl{C}}(\sigma) -S_{\vec\p C}(\sigma) = S_{\hC\setminus C}(\sigma)$ (that is, the boundary contributions simply get a minus sign for an opposite linear curvature of the boundary). Then, we would obtain
\beq\label{ZZ2}
	\Delta_{w\,|\,\p C\cup \p D}^{[\hC_w]} \log Z(C|D) = \frc{c}{48\pi} \Delta_{w\,|\,\sigma}^{[\hC_w]}
		\lt[S_{\cl{C}}(\sigma^\sharp) + S_{\hC\setminus C}(\sigma)\rt]_{\sigma=0}
		= \Delta_{w\,|\,\p C}^{[\hC_w]} \log (Z_C^R Z_{\hC\setminus \cl{C}}^R).
\eeq
On the right-hand side, we have not a single partition function, but a product. Again, this product guarantees that the derivative in directions of small global conformal transformations is zero. Yet, there is no ambiguity as to ``where'' the stress-energy tensor is inserted: the point $w$ must lie in $C$, and the analytic continuation of the function of $w$ that is obtained does not reproduce the derivative at points $w$ outside $C$.

But let us come back to (\ref{ZZ}). Evaluating it directly would need a more precise understanding of the boundary terms in the Liouville actions. However, there is way of relating these boundary contributions to the stress-energy tensor without an explicit evaluation. Indeed, the stress-energy tensor may in fact be defined as the field generating the variation of the partition function under a change of metric $\eta \mapsto \eta+\delta \eta$ \cite{F84}:
\beq\label{basicform}
	\delta \log Z_A = \frc12 \int_{\cl{A}} d^2x\, \bra \delta\eta_{ab}(x) T^{ab}(x)\ket_A.
\eeq
Here, $A$ is some domain, and $T^{ab}$ is the symmetric stress-energy tensor in the canonical normalization (in this normalization, the charge $\int dx\, T^{0a}(x,y)$, in the quantization on the line, generates $x^a$-derivatives with coefficient 1). With tracelessness $T^a_a=0$, it is related to the holomorphic and anti-holomorphic components $T$ and $\b{T}$ via
\beq\label{Txx}
	T = -2\pi T_{zz} = -\pi (T_{xx} - i T_{xy}),\quad \b{T} = 2\pi T_{\b{z}\b{z}} = \pi (T_{xx} + i T_{xy}).
\eeq
This involves both a ``change of coordinates'' $z=x+iy,\,\b{z} = x-iy$, as well as a change of normalization in order to guarantee the correct CFT normalization of $T$ and $\b{T}$.

Under a transformation $g = \id + h$ that is conformal on the domain of definition, with $h$ small, the metric changes diagonally, $\delta\eta_{ab} = (\p h + \b\p \b{h}) \delta_{ab}$, so that we obtain the one-point function of the trace of the stress-energy tensor in (\ref{basicform}). This trace is zero except at the boundary, hence we are left with a boundary integration, as expected by the previous considerations. If we take $h(z) = \frc{\ep}{w-z}$ for some small complex $\ep$, we can evaluate $\Delta_w^{\hC_w} \log Z_{A}$ by extracting the part proportional to $\ep$ in $\delta \log Z_{A}$, and discarding the part proportional to $\b\ep$, as long as $w\not\in A$. If $w\in A$, we have to find a function $h^\sharp$ that has the same infinitesimal effect on $\p A$ but that is holomorphic on $A$. In this way, we could evaluate both terms on the right-hand side of (\ref{ZZ}): the first term by evaluating $\delta \log Z_C$ under $h^\sharp$, the second by evaluating $\delta \log Z_{C\setminus \cl{N(w)}}$ under $h$ and discarding the part that is integrated along $\p N(w)$.

Finding $h^\sharp$ in general is complicated. The simplest way to evaluate $\delta \log Z_C$ under $h^\sharp$ is rather to evaluate $\delta \log Z_{C\setminus \cl{N(w)}}$ under $h$ and take the limit where $N(w)\to \emptyset$ -- we just make a puncture at $w$. Evaluating the contribution of the puncture can be done via (\ref{basicform}), where the bulk metric change $\delta \eta_{ab}$ is singular at $w$, and not diagonal there. Denoting this contribution by $\delta \log Z_C[\mbox{puncture}]$, we simply find that
\[
	\frc{c}{48\pi}\lt.\Delta_{w\,|\,\sigma}^{[\hC_w]} S_{\vec\p C}(\sigma^\sharp)\rt|_{\sigma=0}
		= \frc{c}{48\pi}\lt.\Delta_{w\,|\,\sigma}^{[\hC_w]} S_{\vec\p C}(\sigma)\rt|_{\sigma=0} + \delta \log Z_C[\mbox{puncture}]
\]
and hence that
\beq\label{ZZ3}
	\Delta_{w\,|\,\p C\cup \p D}^{[\hC_w]} \log Z(C|D) = \delta\log  Z_C[\mbox{puncture}].
\eeq
This formula quite directly leads to the one-point function of the stress-energy tensor (see below). In terms of the expression (\ref{ZZ2}), these considerations suggest that the product $Z_C^R Z_{\hC\setminus \cl{C}}^R$ takes care of the boundary conditions, upon inserting the bulk stress-energy tensor, by a ``method of images.'' Also, we see that the presence of the domain $D$ in the relative partition function $Z(C|D)$ has the important effect of cancelling the boundary contributions to the singular metric change, so that only the puncture contribution remains.

The calculation of $\delta Z_C[\mbox{puncture}]$ goes as follows. In general, for a transformation of coordinates $\delta x^a = v^a(x,y)$, the metric change is $\delta \eta_{ab} = \p_a v_b + \p_b v_a$. In our case, we simply have $\delta z = h(z)$, so that
\[
	\p_x v_x + \p_y v_y = \p h + \b\p \b{h},\quad \p_x v_x - \p_y v_y = \b\p h + \p \b{h},\quad
		\p_x v_y + \p_y v_x = -i(\b\p h - \p \b{h}).
\]
Using the formulae \cite{F84}
\[
	\frc{\p}{\p z} \frc{1}{w-z} = \frc{\p}{\p \b{z}} \frc{1}{\b{w}-\b{z}} = -\pi \delta^2(z-w)
\]
it is straightforward to arrive at
\[
	\delta \eta_{ab} T^{ab} = -2\pi \delta^2(z-w) \lt( (\ep+\b\ep) T_{xx} - i (\ep-\b\ep) T_{xy} \rt).
\]
Hence, using (\ref{Txx}) and (\ref{basicform}) and keeping the $\ep$ part only we obtain (\ref{onept}).

\end{document}